\documentclass[aps,pre,twocolumn,superscriptaddress,floatfix]{revtex4-2}

\usepackage[utf8]{inputenc}
\usepackage[T1]{fontenc}
\usepackage[pdftex]{graphicx}
\usepackage{xspace}
\usepackage{xcolor}
\usepackage{amsmath}
\usepackage{amssymb}
\usepackage{gensymb}
\usepackage[thinc]{esdiff}
\usepackage{float}
\usepackage{placeins}
\usepackage{subfigure}
\usepackage{caption}
\usepackage{hyperref}
\hypersetup{hidelinks}

\newcommand{\We}{\ensuremath{{\rm We}}}
\newcommand{\Oh}{\ensuremath{{\rm Oh}}}
\newcommand{\Rey}{\ensuremath{{\rm Re}}}
\newcommand{\De}{\ensuremath{{\rm De}}}
\newcommand{\Ek}{\ensuremath{E_{\rm k}}}
\newcommand{\Et}{\ensuremath{E_{\rm t}}}

\begin{document}

\title{Impact of viscoelastic polymer solution droplets on a granular bed}

\author{Jooyeon Park}
\email[]{jooypark@umd.edu}
\affiliation{Department of Mechanical Engineering, University of Maryland, College Park, Maryland 20742, USA}

\author{Th\'eophile Meiller}
\affiliation{Department of Mechanical Engineering, University of California, Santa Barbara, California 93106, USA}

\author{Sreeram Rajesh}
\affiliation{Department of Mechanical Engineering, University of California, Santa Barbara, California 93106, USA}

\author{Alban Sauret}
\email[]{asauret@umd.edu}
\homepage[]{https://sauretlab.umd.edu/}
\affiliation{Department of Mechanical Engineering, University of Maryland, College Park, Maryland 20742, USA}
\affiliation{Department of Chemical and Biomolecular Engineering, University of Maryland, College Park, Maryland 20742, USA}

\begin{abstract}
The impact of polymer solution droplets on granular beds is relevant to powder processing, binder jetting additive manufacturing, and environmental applications involving erosion control or spray deposition, yet most controlled studies of drop--grain interactions have focused on Newtonian liquids. In this study, we experimentally investigate the impact of viscoelastic polyethylene oxide (PEO) droplets on a dry granular bed and compare the resulting cratering dynamics with those of Newtonian liquids over a wide range of impact energies and Ohnesorge numbers. Crater morphology changes with impact energy, and this evolution occurs at lower energies for drops of polymer solution, consistent with their distinct liquid--grain interactions during impact. The crater diameter exhibits two distinct regimes: a low-energy plateau and a power-law growth at higher impact energies. We identify the transition between these regimes and show that, although the plateau size and the power law remain nearly unchanged, viscoelastic droplets reach the transition at lower impact energy than Newtonian droplets. This suggests that viscoelasticity modifies how the impact energy is partitioned between droplet deformation and dissipation in the granular bed.
\end{abstract}


\maketitle

\section{Introduction}

Droplet impact on granular materials occurs in environmental and industrial processes in which liquid spreading, pore-scale penetration, and grain mobilization are coupled. In natural settings, raindrop impact detaches and redistributes soil grains, contributes to runoff and infrastructure damage, and can generate aerosols and bioaerosols from soil~\cite{Pimentel2006Ecology,Montgomery2007PNAS,Zhao2015PNAS,Prosser2001EarthSurf,Joung2017NatureMicro,joung2017bioaerosol}. Similar drop--grain interactions arise in wet granulation and binder-jet additive manufacturing, where a liquid binder must penetrate the powder bed enough to form bonds while avoiding excessive crater formation, powder ejection, or uncontrolled liquid migration~\cite{Hapgood2002,Marston2010PowderTech,Lawrence2024AddManuf,kumar2025powder}. Related questions also appear in polymer-based soil stabilization, fire-retardant delivery, erosion-control treatments, and agrochemical spraying, where additives are used to tune retention, penetration, and particle cohesion~\cite{Hong2016IBB,markiewicz2024polymeric,Hewitt2024PestManagSci,Yu2019PNAS,jayaprakash2025enhancing}.

The fluid dynamics of droplet impact on rigid and liquid surfaces have been studied extensively~\cite{Worthington1908Drops,Yarin2006AnnRev,josserand2016drop}. Depending on the impact conditions and substrate properties, including more complex targets such as flexible fibers and liquid-infused porous surfaces, outcomes such as spreading, splashing, rebound, crown formation, and cavity development can occur~\cite{Rioboo2001ExpFluids,che2018impact,driscoll2010thin,eggers2010drop,Clanet2004JFM,Dressaire2016SoftMatter,Muschi2018SoftMatter}. These behaviors are commonly characterized using the Weber number, $\We$, the Reynolds number, $\Rey$, and the Ohnesorge number, $\Oh=\mu/\sqrt{\rho\sigma d_0}$, which compare inertia, capillarity, and viscous effects~\cite{Clanet2004JFM,Mobaseri2025PNAS,Wang2022POF,avni2026maximalspreadingimpactingviscoelastic}. When the target is a granular bed, additional processes arise: the liquid penetrates the pore space, grains are mobilized, and the impact energy is partitioned between droplet deformation, imbibition, bed deformation, and particle ejecta~\cite{andreotti2013granular,Zhao2015PNAS,deJong2017PRE,trottet2025sandball}. More generally, liquid redistribution in wet granular materials can control particle capture, aggregation, and cohesion through capillary bridges~\cite{Saingier2017PRL,SharmaSauret2025SoftMatter}. For Newtonian drops on dry granular beds, previous studies have reported crater-size scalings $D_{\rm c}\propto E^{0.17-0.25}$ and impact-energy-dependent crater morphologies~\cite{Katsuragi2010JFM,katsuragi2010morphology,Delon2011SoftMat,Marston2010PowderTech,Zhao2015PNAS}. Viscosity can slow spreading and absorption, but previous measurements suggest that it has a limited effect on the final crater diameter over the range tested so far~\cite{Nefzaoui2012ETFS}. Drop impact on wet granular beds further shows that the liquid content of the bed can modify crater formation and scaling laws~\cite{Zhang2015PRE,zhang2024drop}.

However, many liquids deposited onto powders or soils are formulated fluids rather than simple solvents. In binder jetting and wet granulation, polymeric or viscous binders are used to tune spreading, penetration, and granule strength since excessive spreading can reduce spatial resolution, while excessive grain mobilization can disturb the powder bed.~\cite{Hapgood2002,Marston2010PowderTech,Lawrence2024AddManuf,kumar2025powder} In environmental and agricultural settings, polymer additives are used to improve spray retention, reduce drift, or stabilize granular surfaces against erosion.~\cite{Hewitt2024PestManagSci,Yu2019PNAS,Hong2016IBB,markiewicz2024polymeric,jayaprakash2025enhancing} These applications require control over two coupled processes: the spreading and recoil of the drop, and the degree to which liquid penetration mobilizes grains within the bed. Yet the role of non-Newtonian rheology during drop impact on a granular bed remains largely unexplored. On rigid surfaces, even a small amount of polymer can strongly alter the outcome of the drop impact, for instance, suppressing splashing and promoting filament formation during spreading and pinch-off.~\cite{Bertola2013AdvColInt,Rozhkov2006}
Polymeric liquids can also modify capillary liquid bridges between particles: under dynamic stretching, viscoelastic filaments delay bridge rupture and increase the axial force between grains.~\cite{rajesh2026axial}
These observations suggest that polymer stresses may affect liquid--grain coupling during impact. Despite these insights, how a viscoelastic drop impacting a granular bed interacts with the surface, and how this modifies drop dynamics and crater morphology, remain elusive.

An important question is therefore how polymer additives modify the transfer of momentum and energy from the impacting drop to the granular bed. Here, we investigate the impact of droplets of viscoelastic polymer solutions on a dry granular bed and compare their response with that of Newtonian liquids under similar impact conditions. We find that viscoelasticity does not significantly change either the low-energy plateau in crater diameter or the power-law growth observed at higher impact energies. Instead, it lowers the transition energy between these two regimes. We relate this shift to spreading dynamics and to changes in liquid--grain coupling during impact. Section~\ref{section2:overall} describes the experimental setup and measurement procedures. Section~\ref{section3:overall} presents the crater morphologies and the evolution of crater size with impact energy. Finally, Section~\ref{section4:overall} discusses the physical mechanism underlying the transition.

\section{Experimental methods}\label{section2:overall}

 \begin{figure}[t]
\centering
  \includegraphics[width=0.8\linewidth]{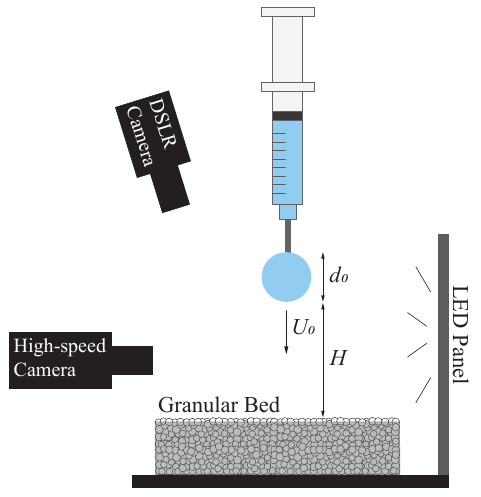}
  \caption{Schematic of the experimental setup.}
  \label{fig:setup}
\end{figure}

The experiment consists of releasing a droplet from various heights onto a cohesionless granular bed, as illustrated in Fig.~\ref{fig:setup}. The droplets were generated by slowly extruding the liquid through a 17G needle (1.5~mm nozzle diameter, McMaster-Carr), yielding a droplet diameter $d_{0} \approx 3-3.6$~mm, depending on the liquid considered. In the following, an overbar denotes normalization by the initial drop diameter $d_0$. The droplet was released from a height $H \in [0.75,\, 60]\,{\rm cm}$ above the granular bed, allowing us to tune the impact velocity $U_0$ and hence the kinetic energy $E_k$. The impact kinetic energy was calculated as $E_k=[m\,{U_0}^2]/2=\pi\,\rho\, {d_0}^3 {U_0}^2/12$, where $m=\rho\pi {d_0}^3/6$ is the mass of the droplet. The drop impact was recorded from the side using a high-speed camera (Phantom VEO 710L, Vision Research) operating at $1000-14000$ fps. After impact, the crater morphology was imaged from above using a DSLR camera (Nikon D5600). In addition, the three-dimensional topography of selected craters was measured using an optical 3D surface profiler (Keyence LJ-S160) to obtain representative surface profiles. Image analysis was performed to quantify the crater diameter (rim-to-rim) using custom MATLAB routines. The diameter was computed as the average of two orthogonal measurements across the crater center to minimize any slight ellipticity. Each data point corresponds to the average of five repeated impacts for each set of conditions.

For cases where the post-recoil absorption stage could be resolved, we also measured the post-recoil imbibition time $t_{\rm imb}$ from the side-view images. We define $t_{\rm imb}$ as the time between the onset of the final post-recoil absorption stage, after recoil or any rebound has ended, and the disappearance of the visible liquid volume above the bed. Therefore, this time corresponds to the late capillary absorption stage, not to the initial pressure-driven penetration during impact.

The granular bed is made of dry glass beads of diameters $78\pm 16\,\mu{\rm m}$ (Potters Industries, see details in Supplementary Materials, Fig. S1). Before each experiment, the granular bed was gently leveled to obtain a reproducible loosely packed configuration with a packing fraction $\phi = 0.59 \pm 0.01$.~\cite{Nefzaoui2012ETFS} In this study, we consider two categories of liquids: Newtonian water--glycerol mixtures and polymer solutions of poly(ethylene oxide) (PEO, $M_w = 4\times10^6~\mathrm{g\,mol^{-1}}$, Sigma-Aldrich). Distilled water (Heidolph Tuttnauer 9000 Water Distiller) and glycerol (Sigma-Aldrich, ACS reagent, $\geq 99.5\%$) were used to prepare the Newtonian water-glycerol mixtures. The water/glycerol mixtures were prepared at various mass ratios (100/0, 75/25, 50/50, 25/75, 10/90, 3/97, and 0/100), resulting in a dynamic viscosity spanning from $\mu = 1~\mathrm{mPa\,s}$ (pure water) to $\mu \sim 10^{3}~\mathrm{mPa\,s}$ (nearly pure glycerol) - measured with an Anton Paar Rheometer MCR302. Over this range, the density varied only weakly, from $\rho \approx 1000$ to $1260~\mathrm{kg\,m^{-3}}$ (measured with an Anton Paar Densimeter DMR 35), and the surface tension remained within $\sigma \approx 63$--$72~\mathrm{mN\,m^{-1}}$ (measured by pendant drop tensiometry). The second set of liquids was viscoelastic solutions prepared by dissolving a small amount of PEO 4M into a 75/25 wt\% water/glycerol solvent mixture at mass fractions of $c=0.05\%$, $0.1\%$, $0.2\%$, $0.3\%$, $0.5\%$, and $1\%$. The PEO solutions were mixed using a roller mixer for 12-24 hours until the polymer was fully dissolved. All polymer solutions were used within 1 week of preparation to minimize degradation or aging of PEO, as its rheological properties can evolve over time.~\cite{rajesh2022transition} $\sigma$ remained approximately constant across all polymer concentrations, with $\sigma \approx 63~\mathrm{mN\,m^{-1}}$. The initial drop diameter $d_0$ and the impact velocity $U_0$ were measured from high-speed imaging for each release height. This direct measurement is especially important for PEO solutions, for which the viscoelastic filament formed during drop detachment can reduce the impact velocity relative to the free-fall value expected from the nominal release height. The droplet diameter $d_0$ ranges from $3.0$ to $3.6$ mm for both Newtonian and PEO solutions and tended to decrease with increasing viscosity and elasticity (detailed information of $d_0$ and $U_0$ is shown in Figure S2).

\begin{figure*}[t]
\centering
        \includegraphics[width=\linewidth]{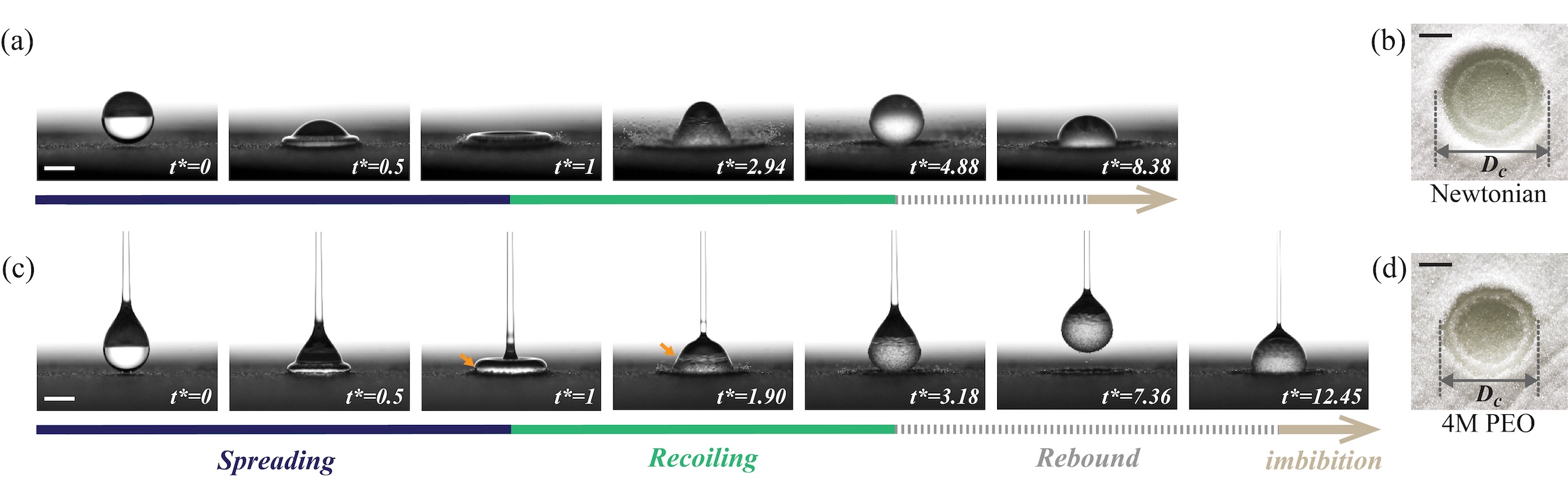}
        \caption{Impact dynamics at low velocity of Newtonian and 4M PEO solution drops on the granular bed. (a) and (c) show the temporal evolution of drop deformation and liquid-grain interaction for the Newtonian drop (W25\%G75\%; $H=2.7~\mathrm{cm}$, $\Oh=0.07$, $\We=27.7$) and the PEO drop (4M PEO 0.5\%; $H=9.7~\mathrm{cm}$, $\Oh=0.07$, $\We=17$), respectively (side-view). (b) and (d) show the corresponding top-view images of the final crater. The crater size, $D_c$, is defined as the outer boundary of the final crater. Time is normalized as $t^*=t/\tau_f$, where $\tau_f$ is the characteristic timescale for maximum spreading. The orange arrows indicate the visible boundary between the upper and lower parts of a lamella for the PEO drop. Scale bars: 2~mm.}
        \label{fig:result1}
    \end{figure*}

The PEO solutions are shear-thinning and elastic. The viscosity $\mu$ was measured with an Anton Paar MCR302 rheometer, and the relaxation time $\lambda$ was obtained from capillary-thinning experiments. These measurements are reported in the Supplementary Material, Fig.~S3. For Newtonian solutions (water--glycerol mixture), the dynamic viscosity is constant and ranges from $1\,\mathrm{mPa\,s}$ to $1410\,\mathrm{mPa\,s}$.

Using the measured values of $d_0$, $U_0$, $\rho$, and $\sigma$, the impact dynamics are characterized by the Weber number, $\We=\rho U_0^2 d_0/\sigma$, the Reynolds number, $\Rey=\rho U_0 d_0/\mu$, and the Ohnesorge number, $\Oh=\mu/\sqrt{\rho\sigma d_0}$. For PEO solutions, the viscosity used in $\Rey$ and $\Oh$ is the effective shear viscosity, $\mu_{\rm eff}$. We evaluate $\mu_{\rm eff}$ at the characteristic shear rate during lamella spreading, $\dot{\gamma}_{\rm c}=U_0/\delta$, where $\delta$ is the effective viscous length scale associated with the spreading lamella. Because $\delta$ depends on $\Rey$, while $\Rey$ depends on $\mu_{\rm eff}$, the quantities $\delta$, $\dot{\gamma}_{\rm c}$, $\mu_{\rm eff}$, and $\Rey$ are obtained self-consistently from the measured viscosity curves, following the framework for impacting generalized-Newtonian drops~\cite{Mobaseri2025PNAS}. Details of the methods are provided in the Supplementary Material, Fig.~S4.

Over the explored conditions, $\We$ spans from $O(10^{-1})$ to $O(10^2)$, $\Rey$ ranges from $O(10^{-1})$ to $O(10^3)$, and $\Oh$ varies between $O(10^{-3})$ and $O(10^1)$. These ranges span both inertia-dominated impacts, $\Oh\ll 1$, and viscous-dominated impacts, $\Oh\gtrsim 1$. Elasticity is characterized separately using the Deborah number, $\De=\lambda/t_{\rm imp}$, with $t_{\rm imp}\sim d_0/U_0$. For the PEO solutions, $\De$ ranges from $O(1)$ to $O(10^2)$, indicating that elastic stresses may influence the impact dynamics.

\begin{figure}[!b]
        \centering
        \includegraphics[width=\linewidth]{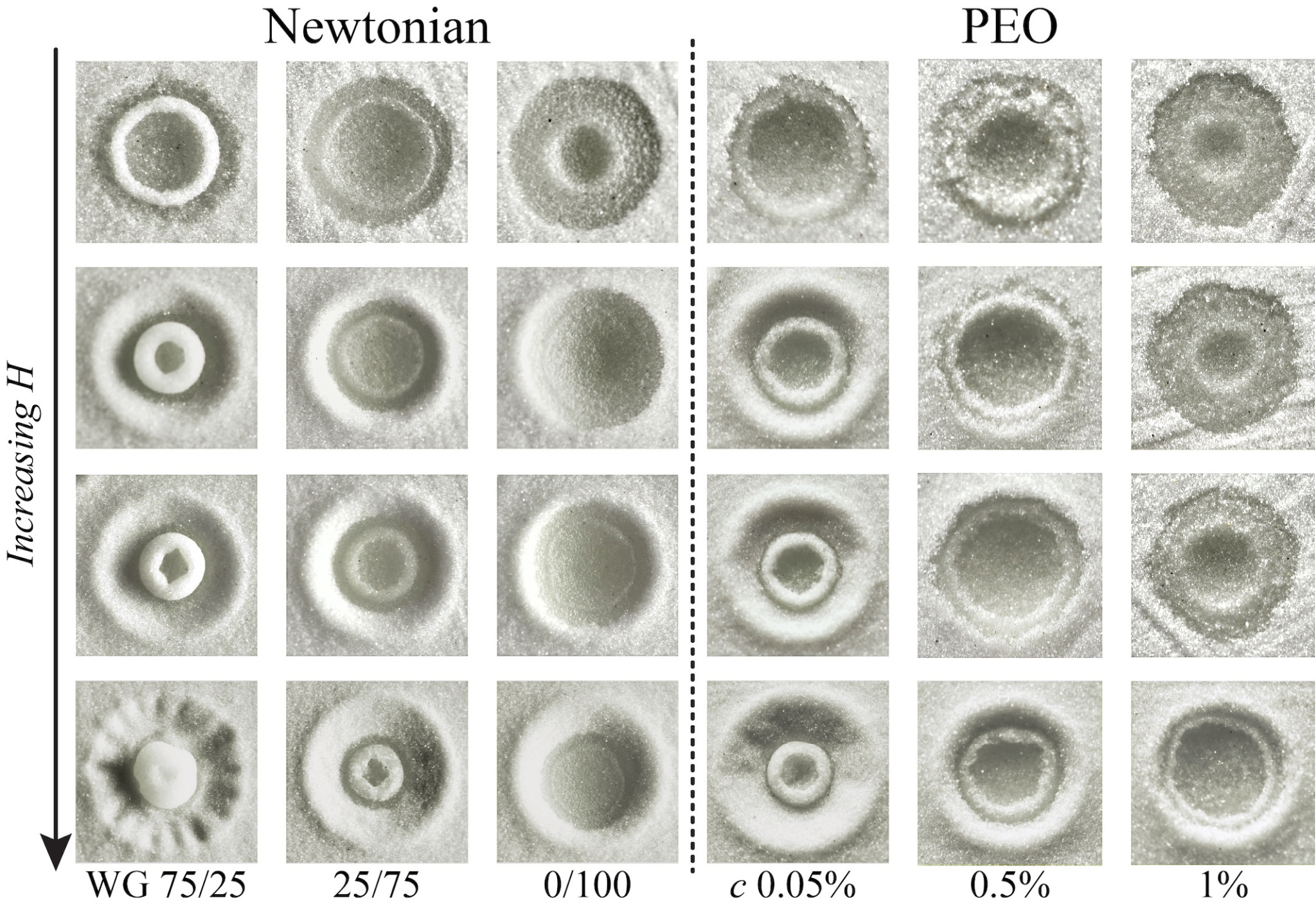}
        \caption{Representative top-view crater morphologies for Newtonian (W75\%G25\%, W25\%G75\%, G100\%) and PEO ($c=0.05\%, 0.5\%, 1\%$) drops over the range of impact conditions explored in this study (not to scale). Release height $H$, and thus impact velocity $U_0$, increases from top to bottom. The images highlight the overall evolution of the crater and the inner morphology with fluid viscosity or elasticity and impact conditions.}
        \label{fig:result2}
    \end{figure}


\section{Results}\label{section3:overall}

\subsection{Impact dynamics and liquid-grain interaction}

Figures ~\ref{fig:result1}(a) and ~\ref{fig:result1}(c) compare representative examples of the impact of Newtonian and of viscoelastic PEO solution on the granular bed at low impact velocity, respectively. Despite the different rheological properties of the two liquids, both droplets exhibit the same overall sequence of events, namely spreading, recoil, rebound, and subsequent static imbibition. This comparison provides a useful starting point for identifying which aspects of the impact response are common to both liquids and which are modified by viscoelasticity.

Although the two cases shown in Figs.~\ref{fig:result1}(a) and~\ref{fig:result1}(c) have comparable Ohnesorge numbers $\Oh$, the Newtonian droplet has a larger $\We$ even though it is released from a height approximately three times lower than the PEO droplet. This difference arises because, at high polymer concentration, the PEO solution forms a long viscoelastic filament during detachment and fall. The filament can keep the drop partially coupled to the needle, thereby reducing the measured impact velocity relative to the free-fall value (Fig.~S2). For this reason, all impact parameters used below are based on the measured values of $U_0$ and $d_0$, rather than on the release height alone.

The post-spreading dynamics also differ. Despite their comparable $\Oh$ and similar surface tension, the PEO droplet completes recoil faster than the Newtonian droplet, with $\Delta t^* \simeq 1.18$ instead of $3.88$, where $t^*=t/\tau_f$ and $\tau_f$ is the characteristic time to maximum spreading. The PEO droplet also rebounds higher. This faster recoil is consistent with elastic stresses generated during filament stretching and deformation of the polymer solution, which can promote contraction after maximum spreading.~\cite{mckinley2002filament,roy2006fall} The larger rebound delays re-contact with the granular bed and postpones the final static-imbibition stage.

A further difference appears immediately after recoil. The Newtonian droplet remains relatively smooth [Fig.~\ref{fig:result1}(a)], whereas the PEO droplet has a rough surface covered with adhered grains [Fig.~\ref{fig:result1}(c)]. Since these two cases have comparable $\Oh$, this contrast is unlikely to arise from viscosity alone. Instead, it suggests reduced grain entrainment into the droplet interior and stronger grain retention at the viscoelastic droplet surface. 

After imbibition, the final craters are characterized by the outer crater boundary and the inner morphology [Figs.~\ref{fig:result1}(b,d)]. We define the crater size $D_c$ as the diameter of the outer boundary, measured rim-to-rim when a raised rim is visible and otherwise from the outer visible edges of the crater or wetted region.

\subsection{Crater morphologies and regimes}
\begin{figure*}
        \centering
        \includegraphics[width=\linewidth]{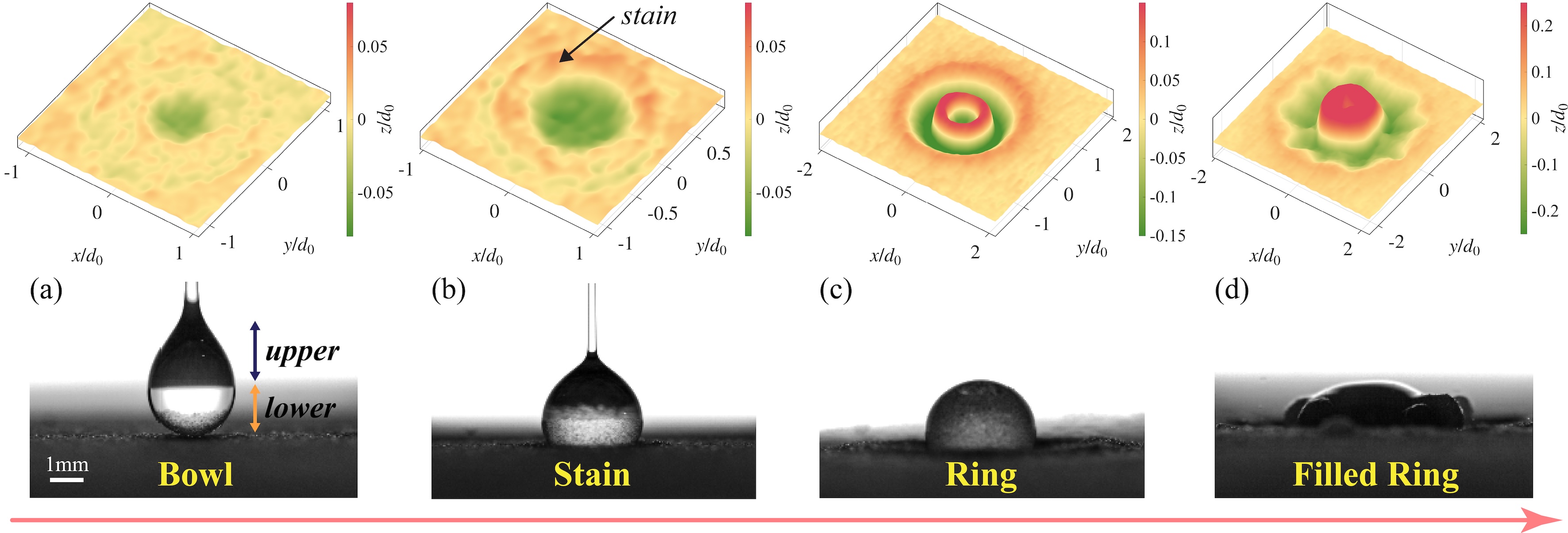}
        \caption{Three-dimensional profiles of representative crater morphologies (top) and corresponding side-view snapshots of drop impact (bottom): Bowl (4M PEO 1\%; $H=7.67~\mathrm{cm}$), Stain (4M PEO 0.5\%; $H=1.16~\mathrm{cm}$), Ring (W75\%G25\%; $H=5.65~\mathrm{cm}$), Filled-ring (W75\%G25\%; $H=34.66~\mathrm{cm}$). The snapshots show the post-recoil droplet, together with the particle-adhered region. The pink arrow denotes the progressive increase in particle mixing with the drop, ranging from surface adhesion to enhanced liquid--particle mixing in the filled-ring case. Scale bar: 1~mm.}
        \label{fig:result3}
    \end{figure*}

\begin{figure*}[t]
        \centering
        \includegraphics[width=0.8\linewidth]{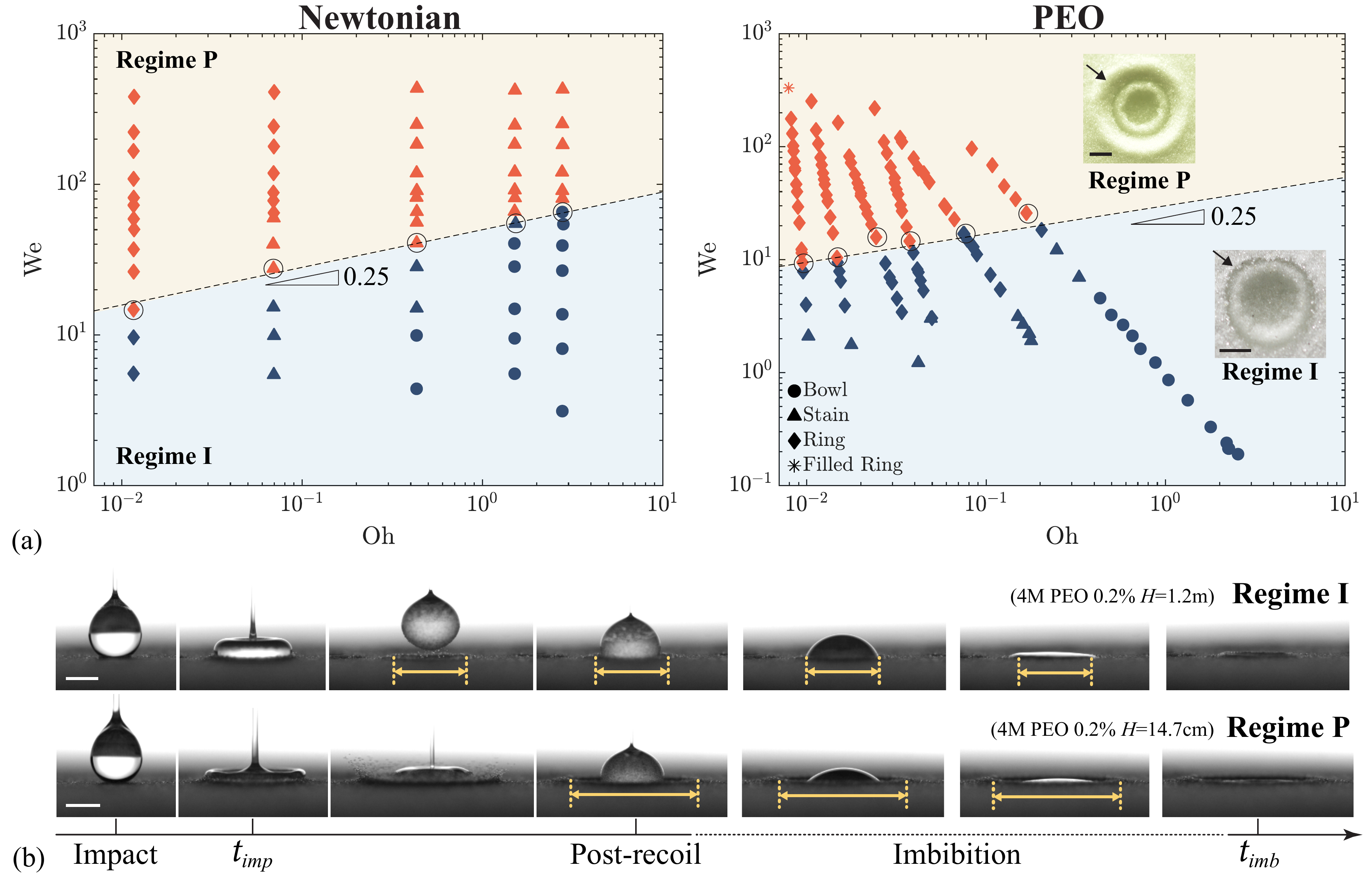}
        \caption{Crater regime maps in the $\We$--$\Oh$ space. (a) Distribution of crater morphologies and the two crater regimes: Regime I, where the final crater boundary is controlled by the imbibed/wetted region, and Regime P, where the final crater boundary is controlled by the particle-ejecta rim. Black circles denote the transition point (between plateau and power-law regions). The insets show crater images for 4M PEO (0.2 wt\%) at $H=1.2~\mathrm{cm}$ and $H=14.7~\mathrm{cm}$, corresponding to Regime I and Regime P, respectively. Blue and orange colors denote Regime I and Regime P, respectively. The dashed lines indicate the empirical transition scalings $\We=50\,\Oh^{0.25}$ for Newtonian liquids and $\We=30\,\Oh^{0.25}$ for PEO solutions. (b) Representative side-view snapshots illustrating the characteristic impact and imbibition dynamics associated with Regime I and Regime P along the timeline from impact to completion of the imbibition. The yellow arrows denote the size of the impact-generated crater. Scale bars: 2~mm.}
        \label{fig:result4}
    \end{figure*} 
Fig.~\ref{fig:result2} provides an overview of the crater morphologies obtained over the range of impact conditions for Newtonian and PEO droplets. For both fluid systems, increasing viscosity or polymer concentration tends to weaken the circular inner morphology, suggesting that fluid rheology modifies the redistribution of liquid and adhered grains during recoil and imbibition. When ring-like morphologies are observed, the inner ring is generally thinner for PEO drops than for Newtonian drops at comparable crater size. These observations motivate the classification introduced in Fig.~\ref{fig:result3}.

Fig.~\ref{fig:result3} shows representative three-dimensional crater morphologies as the impact velocity and thus impact energy increases. As the impact energy increases, the crater morphology systematically evolves from a bowl-type to a stain, a ring, and eventually a filled-ring structure with an asymmetric surrounding crater. Here, the filled-ring morphology denotes a ring-like granular deposit whose center is partially filled, leaving only a shallow central depression rather than an open hole. After recoil, backlighting produces a dark--bright contrast between the upper and lower parts of the droplet when the surface remains relatively free of adhered grains. As grains adhere to the droplet surface, those regions appear darker, progressively obscuring this intrinsic optical contrast. Thus, the morphology labels primarily describe the visible surface distribution of adhered grains after recoil and imbibition, without excluding the entrainment of some grains into the droplet. In the bowl regime [Fig.~\ref{fig:result3}(a)], droplet deformation remains limited, and visible particle adhesion is mostly confined to the lower hemisphere. The resulting crater, therefore, primarily reflects the early contact footprint of the impacting droplet. For highly viscous Newtonian liquids, this early footprint becomes difficult to distinguish at larger release heights because particle ejection and bed deformation smooth the transition between the crater center and the surrounding bed. In contrast, the corresponding bowl morphology in PEO remains shallower and more clearly delineated in top view (compare Fig.~\ref{fig:result3}(a) with Fig.~S8(a)). As shown in Fig.~\ref{fig:result2}, the pure glycerol Newtonian case rapidly evolves into a deeper bowl with a less distinct early-impact footprint as the release height increases, whereas the PEO 1\% case retains a comparatively shallow and well-defined bowl even at larger release heights (see Fig.~S5 for the Newtonian map and Figs.~S6 and S7 for the PEO maps).

As the impact energy increases, particle adhesion extends beyond the hemispherical boundary, indicating increased particle involvement across the drop surface and the evolution toward a circular stain morphology [Fig. \ref{fig:result3}(b)]. At this stage, the stain morphology already differs between the two fluid classes. In the PEO solutions, the stain appears as a broad grain-coated annulus near the outer edge of the bowl. For Newtonian liquids, by contrast, it more often appears as a thinner annulus located inside the bowl. This difference is visible in Fig.~\ref{fig:result2}; for example, the first-row (minimum $H$) cases of W/G 25/75 and PEO 0.5\% already show the distinct stain patterns described here (details in Figs.~S5--S7). With further increase in impact energy, particle adhesion extends over most of the drop surface, and the circular stain evolves into a ring-shaped granular deposit around the crater center [Fig.~\ref{fig:result3}(c)]. At the highest impact energies, the crater on the granular bed exhibits petal-shaped radial undulations rather than a smooth concave shape. This symmetry-breaking state is characterized by strong drop deformation and a marked increase in liquid--grain mixing volume, resulting in a filled-ring morphology [Fig.~\ref{fig:result3}(d)].

The identified morphology types are mapped in the $\We$--$\Oh$ space in Fig.~\ref{fig:result4}(a) for both Newtonian water--glycerol mixtures and PEO solutions. Increasing impact energy (higher $\We$) and decreasing viscosity (lower $\Oh$) promote a morphological evolution from bowl to stain, ring, and eventually filled-ring structures. Additionally, we find two distinct crater regimes based on how the final crater boundary is defined (indicated by the arrows in the inset of Fig.~\ref{fig:result4}(a)). In Regime I, the boundary of the crater is governed by the final wetted area after imbibition. This regime typically occurs at relatively low impact energies, where the impact-generated crater remains smaller than the final wetted region [Fig.~\ref{fig:result4}(b)]. As the impact energy increases, droplet spreading becomes more pronounced, and a substantial particle-ejecta rim forms during impact. When the diameter of this ejecta rim exceeds the final wetted diameter, the rim defines the final crater size. This corresponds to Regime P, as illustrated in Fig.~\ref{fig:result4}(b).

The dividing line between these two regimes follows $\We \sim \Oh^{0.25}$ for both types of liquid, with prefactors of about 50 for Newtonian liquids and 30 for PEO solutions. Thus, PEO drops enter Regime P at lower $\We$. Along this dividing line, Newtonian cases predominantly show stain morphologies, whereas PEO cases already show ring structures. Since PEO drops have lower measured impact velocities, this earlier ring formation is unlikely to result from inertia alone. It is instead consistent with polymer elasticity modifying the redistribution of adhered grains during recoil: in Fig.~\ref{fig:result1}(c,d), grains are transported along the retracting PEO lamella and remain near the surface, whereas the Newtonian case leaves a thinner stain within the bowl despite its larger $\We$.


\subsection{Plateau-to-power-law transition in crater growth}

\begin{figure*}[t]
        \centering
        \includegraphics[width=\linewidth]{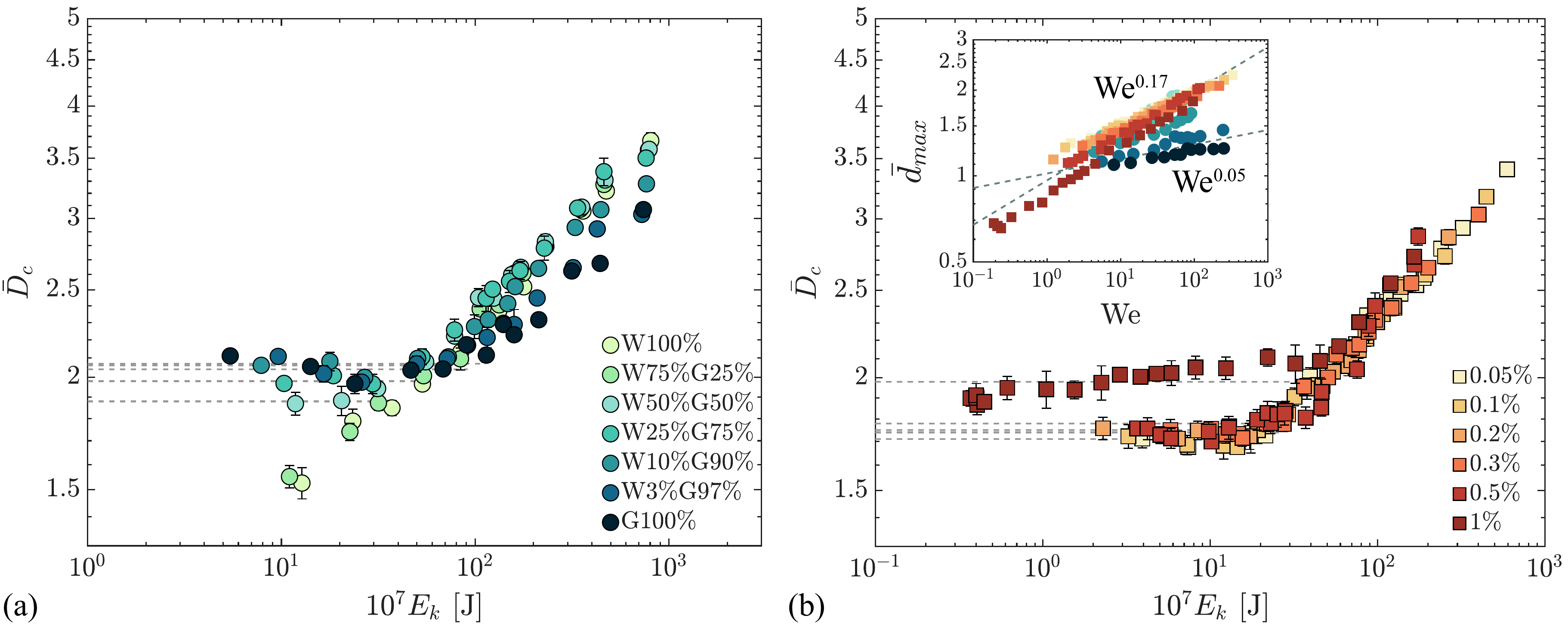}
\caption{Final crater diameter normalized by the initial drop diameter, $\bar{D}_c = D_c/d_0$, as a function of the scaled impact kinetic energy, $10^{7}\,E_k$ (with $E_k$ in joules), for (a) Newtonian solutions and (b) PEO solutions. Gray dashed lines indicate the averaged $\bar{D}_c$ in the plateau regime. Inset: normalized maximum drop spreading diameter ($\bar{d}_{\max} = d_{\max}/d_0$) as a function of $\We$.}
        \label{fig:result5}
    \end{figure*}
    
In this section, we develop a quantitative description of craters by analyzing how impact energy determines crater size. Fig.~\ref{fig:result5} illustrates the evolution of the normalized crater size $\bar{D}_{\rm c}$ as a function of impact kinetic energy $E_{\rm k}$. In the low-energy range, $\bar{D}_c$ exhibits an energy-independent behavior, defining a plateau regime. Both Newtonian and PEO cases collapse within $\bar{D}_c = 1.875 \pm 0.125$. This weak dependence on impact energy suggests that, in the plateau regime, crater size is primarily constrained by geometric and mass-conservation considerations. As a first estimate, we assume that the drop volume fills the pore space of an approximately hemispherical wetted region of diameter $D_c$. Since the bed packing fraction is $\phi=0.59$, the pore volume fraction is $1-\phi$. Equating the drop volume, $\pi {d_0}^3/6$, with the pore volume, $(1-\phi)\pi {D_c}^3/12$, gives
$D_c/d_0 = \left(2/(1-\phi)\right)^{1/3} \simeq 1.7$ (see Supplementary Materials, Fig.~S9(a)). This estimate predicts a crater size consistent with the measured plateau value within 20\%. Furthermore, the plateau value exhibits little dependence on $\Oh$, suggesting that viscous effects are not the primary factor setting the crater size in this low-energy regime. The plateau persists up to a transition energy, marking the onset of an increase of $\bar{D}_c$ with $E_k$, similar to usual observations for Newtonian fluids.

For $E_k > E_t$, $\bar{D}_c$ exhibits a power-law dependence on $E_k$, $\bar{D}_{\rm c} \propto E_k^\alpha$, with an exponent of $\alpha = 0.194 \pm 0.015$. No clear dependence of the exponent on $\Oh$ is observed (Fig.~S9(b)). Previous studies have reported crater scaling exponents ranging from $0.17$ to $0.25$. An exponent near $0.25$ corresponds to spreading-dominated behavior, consistent with the maximum droplet spreading diameter, $\bar{d}_{\max}$, scaling as $\We^{1/4}$ on smooth surfaces. In this case, the final crater size reflects the maximum extent of droplet spreading.\cite{Delon2011SoftMat,Katsuragi2010JFM,Pontier2025SM} Smaller values around 0.17 arise when the impact energy is partitioned between droplet spreading, particle ejection, and bed deformation.~\cite{Zhao2015PNAS} 

We emphasize that this plateau concerns the final crater diameter $\bar{D}_c$, not the transient maximum spreading diameter $\bar{d}_{\max}$. The latter continues to increase with $\We$, as expected for impact-driven spreading, but in the plateau regime, this transient spreading does not set the final crater size after recoil and imbibition. In our experiments, we observe $\bar{d}_{\max} \sim \We^{0.17}$ for $\Oh<1$ in both Newtonian fluids and PEO solutions, as shown in the inset of Fig.~\ref{fig:result5}(b), indicating that energy dissipation through grain interactions already occurs during droplet spreading. In this context, the measured crater-growth exponent of $0.194$ is consistent with energy-partition-controlled behavior. For the high-$\Oh$ Newtonian liquids, where $\bar{d}_{\max}$ becomes only weakly dependent on $\We$ as $\We^{0.05}$, the crater exponent remains nearly unchanged. This suggests that, in this regime, crater growth is not directly governed by the maximum droplet spreading diameter $\bar{d}_{\max}$.


\subsection{Transition impact energy}

    \begin{figure}[!b]
        \centering
        \includegraphics[width=\linewidth]{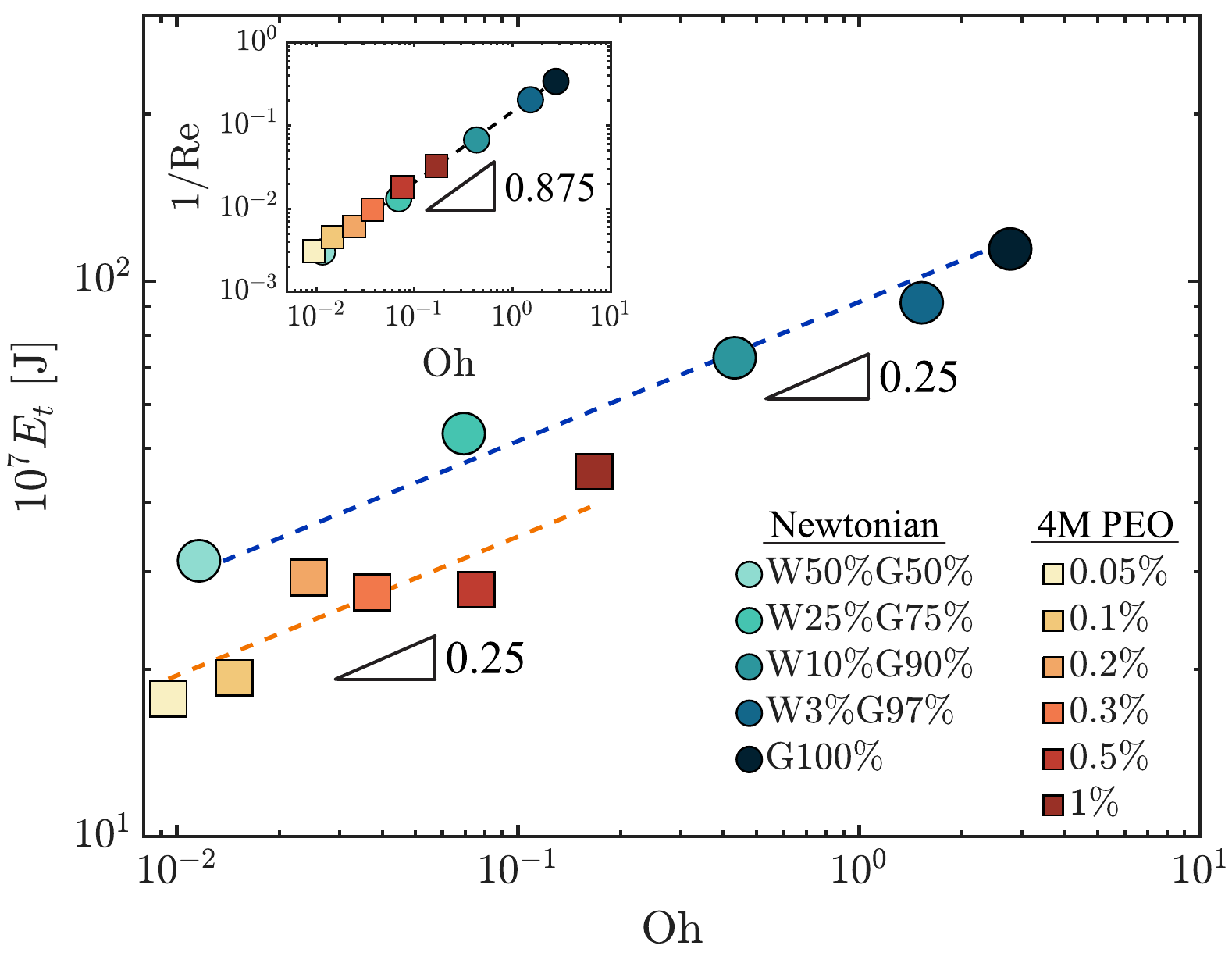}
        \caption{Transition impact energy, $10^{7}\,E_t$ (with $E_t$ in joules), as a function of $\Oh$ for Newtonian and PEO solutions. Inset: Relation between $1/\Rey$ and $\Oh$ at the transition point.}
        \label{fig:result6}
    \end{figure}
The transition energy $\Et$ is defined as the crossover between the low-energy plateau and the power-law growth regime. For each liquid composition, we estimate the plateau value of $\bar{D}_c$ from the low-energy data and fit the high-energy data to $\bar{D}_c=A E_k^\alpha$. The transition energy is then defined as the intersection between these two fits. This procedure is independent of the visual crater classification, but the resulting transition points are consistent with the onset of the rim-controlled Regime P.

Fig.~\ref{fig:result6} shows that $\Et \sim \Oh^{0.25}$ for both Newtonian and PEO solutions. This scaling follows from the empirical transition relation $1/\Rey \sim \Oh^{0.875}$ and the identity $\Oh=\sqrt{\We}/\Rey$, which gives $\We \sim \Oh^{0.25}$. Since $d_0$ and $\sigma$ vary only weakly across the fluids, $\We$ is proportional to $\Et$ in our experiments. The exponent matches the crater-regime dividing line in Fig.~\ref{fig:result4}, and the transition points lie close to that line. However, the prefactor is lower for PEO solutions, so for a given $\Oh$, PEO drops enter the rim-controlled regime at a lower transition energy.

\begin{figure*} [t]
        \vspace*{2.5mm}
        \centering
        \includegraphics[width=\linewidth]{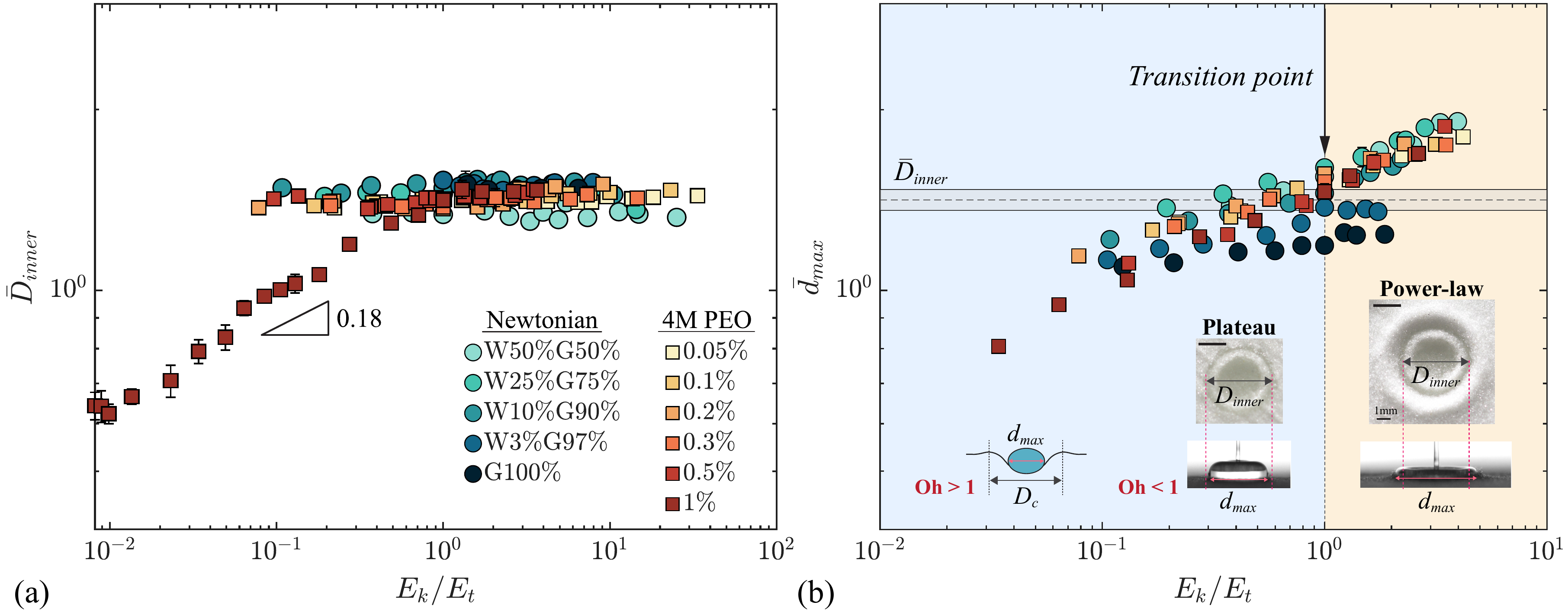}
        \caption{(a) Crater inner morphology diameter normalized by initial drop diameter, $\bar{D}_{\rm inner}=D_{inner}/d_0$, as a function of normalized impact energy $E_k/E_t$, where $E_k/E_t=1$ denotes the transition point. For a given fluid, increasing $E_k/E_t$ corresponds to increasing release height. (b) Maximum spreading diameter normalized by initial drop diameter $\bar{d}_{\max}=d_{max}/d_0$ as a function of $E_k/E_t$. The horizontal dashed line in a shaded band denotes the mean value of $\bar{D}_{\rm inner}$, and the band represents $\pm$ one standard deviation. The insets show representative plateau (4M PEO 0.2\%; $H=0.95~\mathrm{cm}$) and power-law (4M PEO 0.2\%; $H=9.7~\mathrm{cm}$) cases for $\Oh<1$, and also a schematic illustration of the low-spreading behavior for $\Oh>1$. Scale bars: 2~mm.}
        \label{fig:discussion1}
    \end{figure*}


\section{Discussion}\label{section4:overall}
   
We now discuss the mechanism setting the transition energy and why PEO solutions reach this transition at lower energy than Newtonian liquids at the same $\Oh$.

\subsection{Physical mechanism of the transition}
Since the transition marks the point at which the crater boundary shifts from the imbibed wetted region to the particle-ejecta rim, it suggests a competition between two characteristic length scales: the transient maximum spreading diameter, $\bar{d}_{\max}=\max_t[d(t)]/d_0$, reached at $t=\tau_f$ before recoil, and the inner morphology size $\bar{D}_{\rm inner}$. Here, $\bar{D}_{\rm inner}=D_{\rm inner}/d_0$ denotes the diameter of the visible inner morphology. For bowl and stain morphologies, $D_{\rm inner}$ is measured across the outer boundary of the visible inner deposit or depression. For ring and filled-ring morphologies, the measurement is taken from the outer edge of the annulus on one side to the outer edge on the opposite side; it is not the ring width. This length scale is set by particle rearrangement within the drop during static imbibition after recoil (Fig.~\ref{fig:discussion1}).
   
As shown in Fig.~\ref{fig:discussion1}(a), $\bar{D}_{\rm inner}$ converges to a nearly constant value for most cases. This behavior is consistent with the observation that the plateau crater size (reported in Fig.~\ref{fig:result5}), defined by the imbibition-wetted region, remains nearly independent of $\Ek$. That is, the saturation value is primarily governed by the drop size and the characteristics of the granular bed, which determines the effective amount of particles incorporated into the particle--liquid mixture remaining after impact.~\cite{Zhao2015PNAS} We note that $\bar{D}_{\rm inner}$ is not reported for the high $\Oh$ ($\Oh >1$) Newtonian bowl-type cases because the bowl boundary becomes ambiguous due to the gradual slope created by particle ejecta.

The 1\% PEO solution is the main exception: its bowl-like inner morphology remains well defined and increases as $\bar{D}_{\rm inner}\sim (E_k/E_t)^{0.18}$ [Fig.~\ref{fig:discussion1}(a)]. Because the detachment filament lowers the measured impact velocity, these data sample lower impact energies where the bowl reflects the initial spreading footprint rather than an imbibition-controlled size.

Returning to the energy-independent behavior of $\bar{D}_{\rm inner}$, we observe that $\bar{d}_{\max}$ increases with impact inertia, with $\bar{d}_{\max}\sim \We^{0.17}$ for $\Oh<1$, as shown in the inset of Fig.~\ref{fig:result5}(b). The transition approximately coincides with the point where $\bar{d}_{\max}$ begins to exceed the saturated $\bar{D}_{\rm inner}$. This indicates that once the spreading drop size exceeds the imbibition-controlled region, the initial particle-ejecta rim is preserved and subsequently grows as $\bar{d}_{\max}$ increases, consistent with the mechanism described for Regime P earlier. In contrast, for $\Oh>1$, droplet spreading is strongly suppressed ($\bar{d}_{\max}$ is roughly constant and a best fit to our experimental results gives $\bar{d}_{\max} \propto \We^{0.05}$), and the condition $\bar{d}_{\max} > \bar{D}_{\rm inner}$ is not satisfied even in the power-law regime. Nevertheless, the crater growth exponent remains comparable to that observed for $\Oh<1$. This suggests that in this regime, crater formation is governed less by lateral spreading and more by localized axial deformation of the granular bed, accompanied by the rapid generation of particle ejecta. This behavior is illustrated schematically in the inset of Fig.~\ref{fig:discussion1}(b), and representative impact dynamics showing the weak lateral spreading and axial bed deformation are provided in Fig.~S8(a) in the Supplementary Materials. Importantly, the particle-ejecta--driven growth in the power-law regime is similar for both cases.

\begin{figure}[t]
        \centering
        \includegraphics[width=\linewidth]{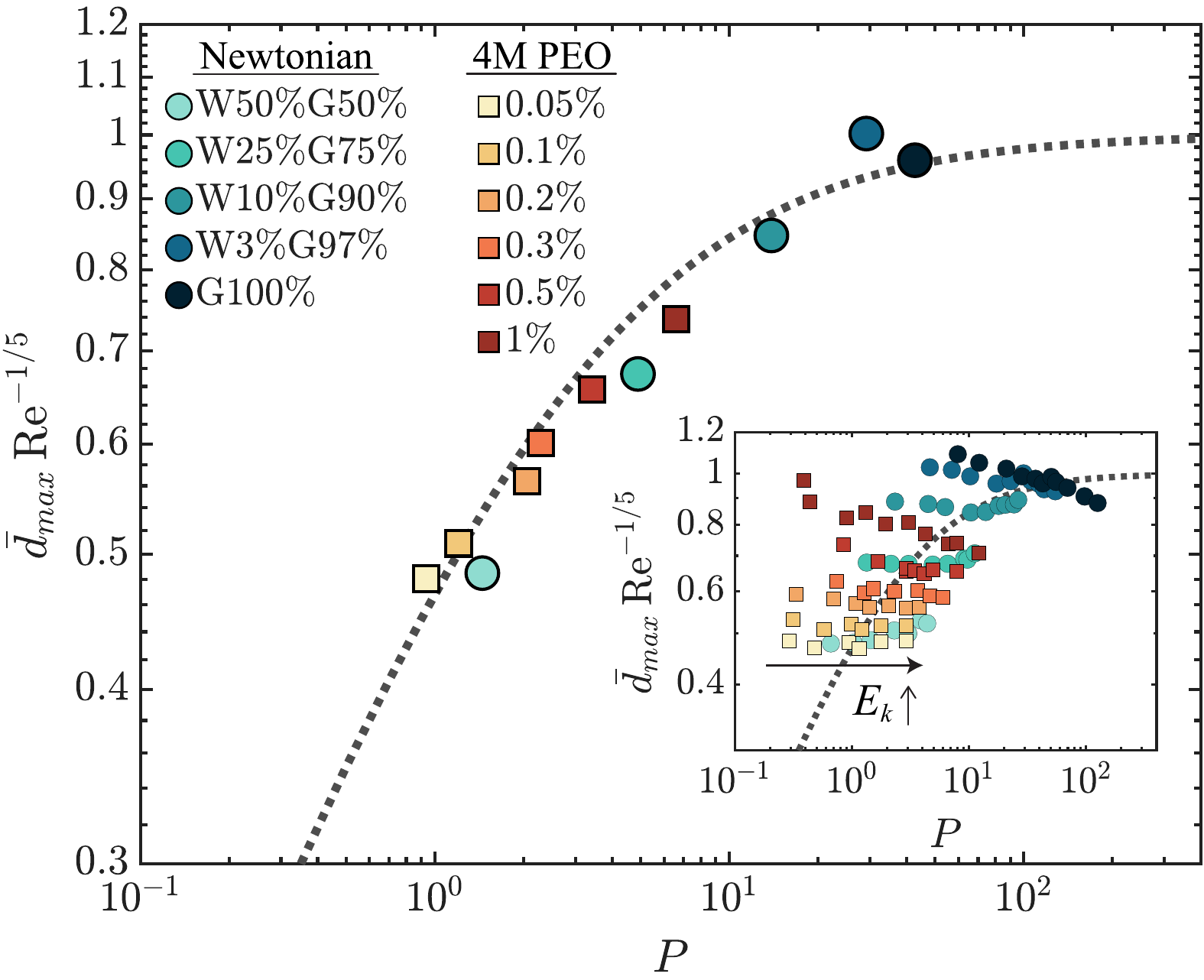}
        \caption{The rescaled maximum spreading diameter, $\bar{d}_{\max} \Rey^{-1/5}$, as a function of the impact number $P = \We \Rey^{-2/5}$ at the transition points for all fluids. The deep-gray dotted line denotes the second-order Pad\'e approximant (Eq.~\ref{eq:pade}). The inset shows the full dataset in the same representation, illustrating the evolution with increasing impact kinetic energy $E_k$.}
        \label{fig:discussion2}
    \end{figure}

Fig.~\ref{fig:discussion2} compares the measured maximum spreading with the smooth-surface scaling written in terms of the impact number $P=\We\Rey^{-2/5}$. The full granular-bed dataset does not collapse onto the smooth-surface prediction (inset of Fig.~\ref{fig:discussion2}), but the transition points for both Newtonian and PEO drops fall close to the Pad\'e approximation,
\begin{equation}
\bar{d}_{\max}\Rey^{-1/5}  = \frac{P^{1/2} + 0.76P}{2 + P^{1/2} + 0.76P}.
\label{eq:pade}
\end{equation}
Before the transition, spreading slightly exceeds the smooth-surface prediction, possibly because the granular material reduces the effective resistance to lamella motion. Beyond the transition, spreading falls below the prediction, consistent with increased energy transfer to bed deformation, particle ejecta, and imbibition. The transition, therefore, corresponds to the point where these spreading-promoting and spreading-limiting effects approximately balance.
    
\subsection{Impact kinetic energy partition}
\begin{figure*}
\centerline{\includegraphics[width=0.8\linewidth]{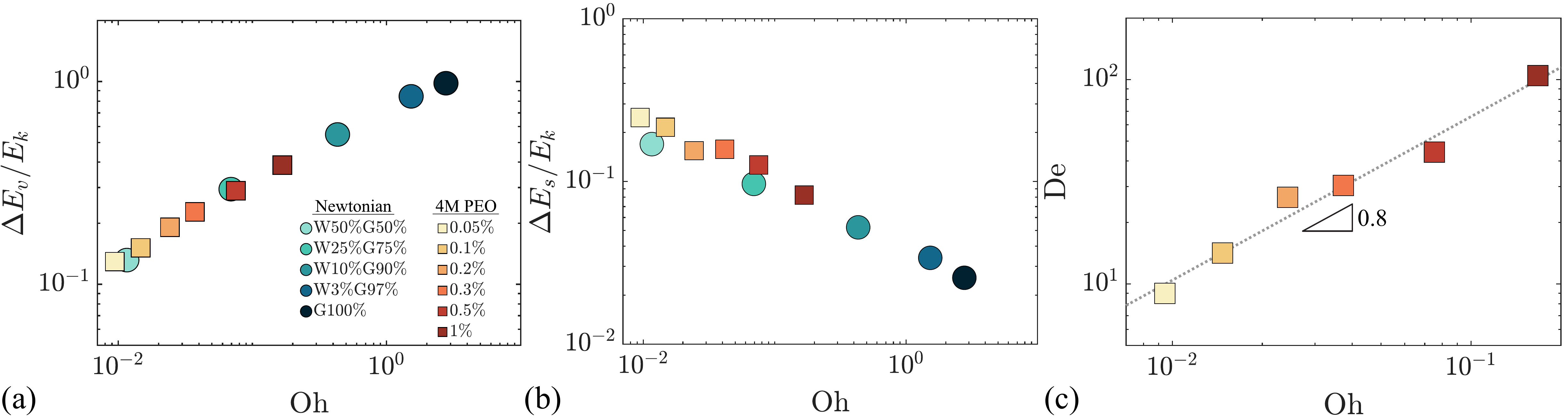}}
        \caption{Energy partition at the transition point as a function of $\Oh$. (a) Viscous dissipation relative to the impact kinetic energy, $\Delta E_{\rm v}/\Et$. (b) Surface energy contribution, $\Delta E_{\rm s}/\Et$. (c) Deborah number $\De$ at the transition point as a function of $\Oh$.}
        \label{fig:discussion4}
    \end{figure*}

As shown in Fig.~\ref{fig:discussion1}(b), the maximum spreading diameter at the transition point is nearly independent of $\Oh$ and fluid type. A direct plot of the transition values further confirms that $\bar{d}_{\max}$ converges to approximately 1.5 for all fluids (see Supplementary Materials, Fig.~S10). Therefore, the lower $\Et$ of PEO solutions is not due to enhanced maximum spreading, but rather points to differences in impact-energy partitioning. To discuss the origin of the reduced transition kinetic energy in PEO solutions, we consider a simplified scaling balance at maximal spreading:
\begin{equation}
E_{\rm k} \sim \Delta E_{\rm s} + \Delta E_{\rm v} + \Delta E_\lambda + E_{\rm bed},
\label{eq:energy}
\end{equation}
where $\Delta E_{\rm s}$ denotes the change in surface energy, $\Delta E_{\rm v}$ the viscous dissipation within the drop, 
 $\Delta E_\lambda$ the elastic energy temporarily stored in the viscoelastic fluid and $E_{\rm bed}$ the bed-related contribution.~\cite{Clanet2004JFM,Mobaseri2025PNAS,avni2026maximalspreadingimpactingviscoelastic} The role of elastic energy in the spreading dynamics is first considered: $\Delta E_\lambda \sim V_{\rm shear}G\gamma^2$, where $V_{\rm shear}$ denotes the effective volume of liquid sheared during lamella spreading, $\gamma$ is the representative strain, and $G$ is the time-dependent elastic modulus of the fluid. This volume is distinct from the stretched detachment filament, which is dominated by extensional deformation and mainly affects drop release and recoil rather than the lamella energy balance at maximum spreading. Avni \textit{et al.} derived, within a Maxwell framework, that the elastic contribution becomes negligible compared to viscous dissipation in the limits $\De\ll1$ and $\De\gg1$ on the smooth surface, and confirmed experimentally that the maximal spreading collapses onto the Newtonian scaling.~\cite{avni2026maximalspreadingimpactingviscoelastic} In the present study, $\De$ exceeds 1 for all PEO solutions and typically reaches $\De \sim O(10)$ at the transition points [Fig.~\ref{fig:discussion4}(c)]. Although these values do not correspond to a strict asymptotic $\De \gg 1$ limit, they place the polymer solutions on the high-$\De$ side relative to the impact time. We therefore do not assume that elasticity is absent; rather, we ask whether elastic energy storage in the spreading lamella controls the maximum spreading. The nearly identical values of $\bar{d}_{\max}$ at the transition for Newtonian and PEO drops suggest that, in the present conditions, lamella-scale elastic storage is not the most important effect, and the maximum spreading remains effectively Newtonian-like even on a granular bed.

Accordingly, the energy variations occurring within the drop at maximal spreading reduce to $\Delta E_{\rm s}$ and $\Delta E_{\rm v}$. We estimate the viscous dissipation in the sheared lamella as $\Delta E_{\rm v}\sim \mu V_{\rm shear}\tau_f\dot{\gamma}^2$, where $V_{\rm shear}\sim {d_{\max}}^2\delta$ is the effective sheared volume and $\mu$ denotes the Newtonian viscosity or the effective shear viscosity $\mu_{\rm eff}$ for PEO solutions. Taking $\tau_f\sim t_{\rm imp}$,~\cite{Clanet2004JFM} the viscous dissipation scales as
\begin{equation}
\Delta E_{\rm v} \sim \mu ({d_{\max}}^2\delta) \frac{d_0}{U_0} \left(\frac{U_0}{\delta}\right)^2 \sim \mu U_0 {d_{\max}}^2 \Rey^{0.5}.
\label{eq:viscousE}
\end{equation}
Here, the boundary-layer thickness is estimated as $\delta \sim d_0\Rey^{-0.5}$, obtained from an inertia-viscous balance using the effective viscosity evaluated at the characteristic shear rate. Surface energy depends on lamella size, which scales with ${d_{\max}}$. This yields:
\begin{equation}
\Delta E_{\rm s} \sim \sigma {d_{\max}}^2.
\label{eq:surfaceE}
\end{equation}
Using Eqs.~\ref{eq:viscousE} and~\ref{eq:surfaceE}, the relative contributions with respect to the initial impact kinetic energy ($E_{\rm k} \sim \rho {d_0}^3 U_0^2$) become:
\begin{equation}
\Delta E_{\rm v}/E_{\rm k}
\sim 
\bar{d}_{\max}^2 \Rey^{-1/2},
\qquad
\Delta E_{\rm s}/E_{\rm k} 
\sim 
\bar{d}_{\max}^2 \We^{-1}.
\label{eq:relativeE}
\end{equation}
These are evaluated at the transition (\textit{i.e.}, $E_k=E_t$), and shown as a function of $\Oh$ in Fig.~\ref{fig:discussion4}(a,b). These quantities should be interpreted as scaling estimates for comparing trends, rather than as exact energy fractions. For Newtonian liquids, increasing $\Oh$ naturally leads to enhanced viscous dissipation during impact, reflected by an increase in $\Delta E_v/E_t$ accompanied by a reduction in $\Delta E_s/E_t$. Importantly, the same trend is observed for the viscoelastic PEO solutions with increasing $\Oh$ (and also $\lambda$),~\cite{Mobaseri2025PNAS} closely overlapping with the Newtonian data. This suggests that, within the accuracy of this scaling estimate, elasticity does not substantially change the relative viscous and surface-energy contributions within the spreading lamella. Meanwhile, since PEO solutions reach the same $\bar{d}_{\max}$ at a lower $\Et$ than Newtonian liquids, the absolutes value of $\Delta E_{\rm v}$ and $\Delta E_{\rm s}$ are correspondingly smaller for PEO solutions.

The lower transition energy of PEO solutions can therefore be interpreted as reflecting a potentially weaker pore-scale coupling between the spreading drop and the granular bed, rather than weaker particle adhesion at the drop surface. In Eq.~\ref{eq:energy}, the bed-related contribution $E_{\rm bed}$ includes bed deformation, $E_{\rm deform}$, and imbibition-associated liquid--grain mixing, $E_{\rm imb}$.~\cite{deJong2021PRF,Delon2011SoftMat,Zhang2015PRE,Zhao2015PNAS,zhao2017liquidgrainmixing} We propose that pressure-driven imbibition and liquid--grain mixing are weaker for PEO solutions than for Newtonian liquids. Since imbibition drains liquid and momentum from the spreading lamella, stronger imbibition in Newtonian liquids acts as an additional energy sink. This interpretation is consistent with previous observations that liquid--grain mixing can suppress droplet spreading and splashing during impact.~\cite{zhao2017liquidgrainmixing} By contrast, if pore-scale penetration is weaker in PEO solutions, the same maximum spreading, and therefore the same rim-controlled transition, can be reached at a lower $E_{\rm t}$.

\begin{figure}[!htbp]
        \centering
        \includegraphics[width=\linewidth]{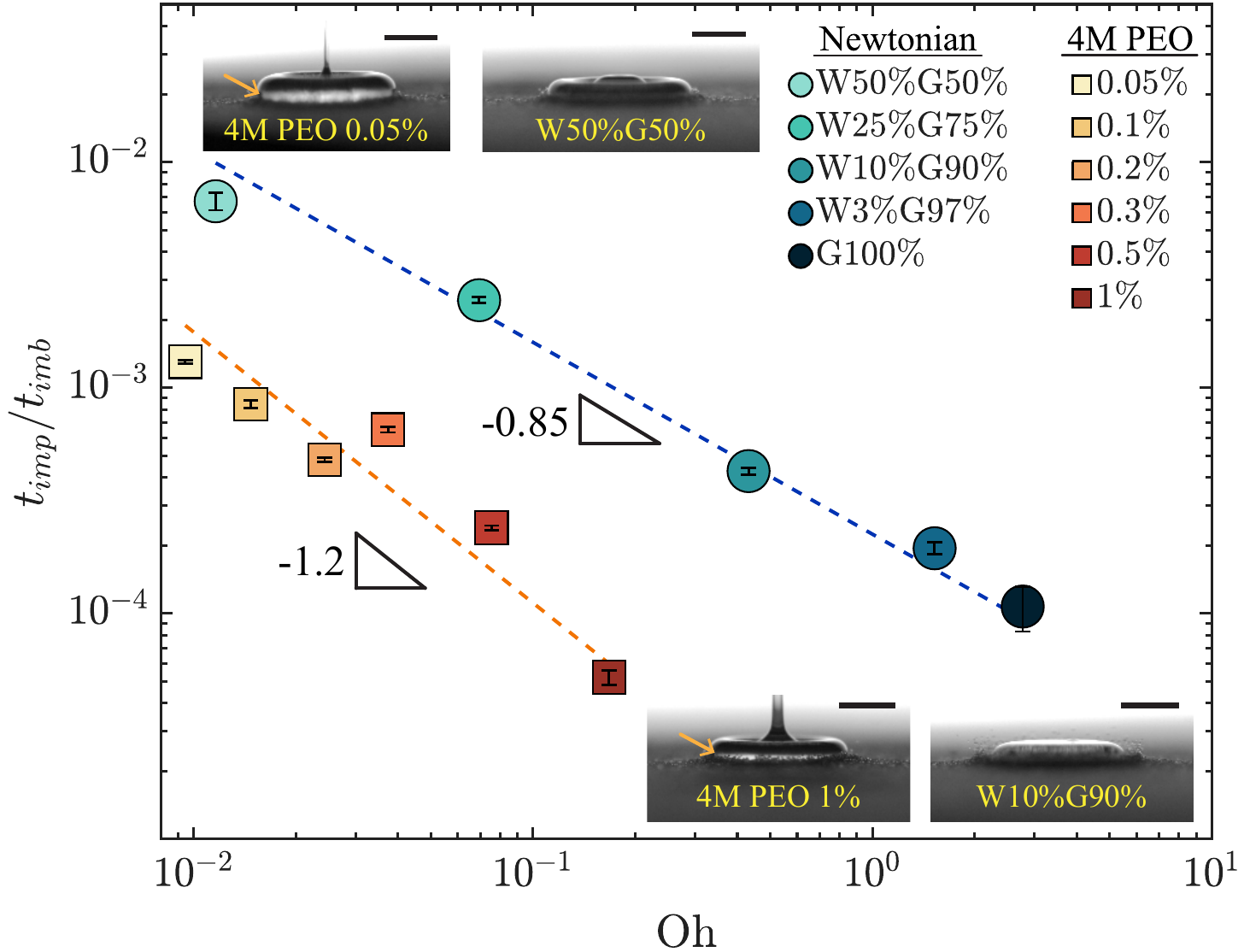}
        \caption{Post-recoil imbibition timescale compared with the impact timescale. Ratio of the impact timescale to the post-recoil imbibition timescale, $t_{\rm imp}/t_{\rm imb}$, as a function of $\Oh$ for Newtonian and PEO solutions at the transition point. Dashed lines indicate power-law fits with exponents -0.85 for Newtonian liquids and -1.2 for PEO solutions. Insets show representative impact images at the transition point for selected cases. The orange arrows denote the bright thin band observed near the lower part of the lamella, indicating less liquid-particle mixing. Scale bars: 2~mm.}
        \label{fig:discussion5}
    \end{figure}

To describe the imbibition of a liquid into a porous medium, we use the Lucas--Washburn framework, in which liquid penetration is driven by capillary pressure and can be modified by an externally imposed impact pressure.~\cite{washburn2021PR} We compare two distinct timescales: the impact timescale, $t_{\rm imp}\sim d_0/U_0$, and the post-recoil imbibition timescale, $t_{\rm imb}$, measured at the transition condition. We find that $t_{\rm imb}$ is two to four orders of magnitude longer than $t_{\rm imp}$ (Fig.~\ref{fig:discussion5}). Thus, late capillary imbibition is unlikely to set the dynamics during the short impact event; instead, liquid penetration during impact is expected to be dominated by the impact-induced pressure.~\cite{Zhao2015PNAS} Because we do not directly resolve the pore-scale flow during impact, the following interpretation should be viewed as a possible mechanism rather than direct evidence. For Newtonian liquids, the pore-scale resistance is primarily set by the bulk viscosity. For polymer solutions, however, confined and extensional pore-scale flows may generate additional viscoelastic stresses and enhanced flow resistance in porous media,~\cite{browne2021elastic} so the resistance to penetration can differ from that inferred from the shear viscosity used to describe spreading. Since $\De=O(10)$ or larger at the transition, such stresses can develop over the impact timescale and may reduce liquid penetration and liquid--grain mixing in PEO solutions. In this context, stronger imbibition provides an additional pathway for transferring impact-driven momentum and energy into the granular bed. If imbibition and liquid--grain mixing are reduced for PEO drops during impact, this pathway is weakened, and less energy is dissipated through the bed.

Indirect support for this interpretation is provided by post-recoil imbibition measurements (Fig. \ref{fig:discussion5}). $t_{\rm imp}/t_{\rm imb}$ follows different empirical trends for the two liquid types: approximately $\Oh^{-1.2}$ for PEO and $\Oh^{-0.85}$ for Newtonian fluids, with the Newtonian data lying systematically above the PEO data. Even under capillary-driven conditions ($\Oh < 0.1$), PEO exhibits significantly slower penetration than Newtonian liquids. Thus, under impact conditions, where larger deformation rates and pressure gradients are imposed, the resistance mechanisms are expected to become more pronounced for PEO solutions. Additionally, the steeper trend observed for PEO is consistent with elastic effects contributing to the suppression of imbibition. Furthermore, this explanation is also consistent with the visual appearance of PEO drops at the transition (see inset of Fig. \ref{fig:discussion5}). A bright thin band is observed near the lower part of the lamella, indicating less liquid-particle mixing. This feature is observed across PEO concentrations at the transition and is consistent with suppressed imbibition. This interpretation suggests that a smaller $E_{\rm bed}$ associated with reduced imbibition for PEO solutions can allow the same spreading performance to be achieved at a lower $E_t$.

\section{Conclusion}

In this study, we experimentally investigated the impact of viscoelastic PEO solution droplets on a dry granular bed made of fine glass beads and compared the results with Newtonian water--glycerol mixtures over a broad range of impact energies and Ohnesorge numbers. The crater morphology evolves systematically with increasing impact energy, from bowl to stain, ring, and filled-ring structures. This evolution occurs at lower impact energies for PEO drops than for Newtonian liquids at comparable $\Oh$, indicating that a small amount of high-molecular-weight polymer can shift the onset of crater-morphology transitions.

The final crater diameter exhibits two regimes: a low-energy plateau and a higher-energy power-law growth regime. The plateau diameter and the power-law exponent are nearly unchanged between Newtonian and PEO drops, with $\bar{D}_c \sim E_k^\alpha$ and $\alpha = 0.194 \pm 0.015$ in the power-law regime. This exponent is consistent with previous studies of Newtonian drop impact on dry granular beds, where crater-size exponents in the range $0.17$--$0.25$ have been reported.~\cite{Katsuragi2010JFM,Delon2011SoftMat,Marston2010PowderTech,Zhao2015PNAS,Pontier2025SM} Thus, over the range of viscoelastic effects considered here, viscoelasticity does not measurably alter this crater-growth exponent within experimental uncertainty. Instead, it lowers the transition energy at which the crater leaves the plateau and becomes rim-controlled.

This shift is not caused by enhanced maximum spreading, since the maximum spreading diameter at the transition remains approximately $\bar{d}_{\max}\simeq 1.5$ for both Newtonian and PEO drops. Rather, the results suggest that polymer additives modify drop--powder interactions beyond an effective-viscosity effect. PEO drops exhibit slower post-recoil imbibition and weaker apparent liquid--grain mixing, suggesting a reduced pathway for transferring impact-driven momentum and energy into the granular bed. This interpretation is consistent with previous observations that liquid--grain mixing can affect spreading and splashing during impact,~\cite{zhao2017liquidgrainmixing} and with the central role of liquid penetration in granule formation and binder performance.~\cite{Hapgood2002,Marston2010PowderTech,kumar2025powder}

More broadly, these results show that the impact of polymer solutions on granular beds is controlled by coupled drop-scale spreading, pore-scale penetration, capillary imbibition, polymer stresses, and grain mobilization. This coupling is relevant to binder-jet additive manufacturing, wet granulation, spray deposition, and erosion-control treatments, where polymeric formulations are used to tune spreading, penetration, and particle cohesion.~\cite{Lawrence2024AddManuf,kumar2025powder,Hewitt2024PestManagSci,Yu2019PNAS,markiewicz2024polymeric} A remaining limitation is that the short-time pore-scale flow during impact is not directly resolved. Future work should measure such fast dynamics, imbibition fronts, and pore-scale extensional flow, while independently varying polymer molecular weight, relaxation time, and solvent viscosity to separate shear-thinning, elasticity, and extensional resistance.

\bigskip\bigskip

\section*{Data availability}
The data supporting this article, including MATLAB scripts, datasets in CSV and XLSX formats, images, and videos, have been deposited in Mendeley Data and are available at https://doi.org/10.17632/kt5wxjptzd.1.

\section*{Conflicts of interest}
There are no conflicts to declare.

\section*{Acknowledgments}

This material is based upon work supported by the Gordon and Betty Moore Foundation GBMF13831 and the National Science Foundation CBET PMP 2526651 and PMP 2533460.

\bibliographystyle{apsrev4-2}
\bibliography{biblio_drop}

\clearpage
\onecolumngrid
\setcounter{figure}{0}
\renewcommand{\thefigure}{S\arabic{figure}}
\setcounter{equation}{0}
\renewcommand{\theequation}{S\arabic{equation}}

\begin{center}
{\Large\bfseries Supporting Material for\\[0.5em]
Impact of viscoelastic polymer solution droplets on a granular bed}

\vspace{1em}
Jooyeon Park$^{1}$, Th\'eophile Meiller$^{2}$, Sreeram Rajesh$^{2}$, and Alban Sauret$^{1,3}$

\vspace{0.5em}
$^{1}$Department of Mechanical Engineering, University of Maryland, College Park, Maryland 20742, USA\\
$^{2}$Department of Mechanical Engineering, University of California, Santa Barbara, California 93106, USA\\
$^{3}$Department of Chemical and Biomolecular Engineering, University of Maryland, College Park, Maryland 20742, USA
\end{center}

\vspace{1em}


\section*{Movies of the experiments}
\begin{itemize}
   \item "Newtonian\_movie1.mp4": Movie corresponding to Figure 2(a) in the main article. Newtonian drop (W25\%G75\%; $H=2.67~\mathrm{cm}$, $\Oh=0.07$, $\We=27.65$), played back 107 times slower than real time.
   \item "PEO\_movie2.mp4": Movie corresponding to Figure 2(c) in the main article. PEO solution drop (4M PEO 0.5\%; $H=9.68~\mathrm{cm}$, $\Oh=0.07$, $\We=16.9$), played back 107 times slower than real time.
\end{itemize}

\vspace{1.5cm}
\section*{Particle distribution in granular bed}

\smallskip

\begin{figure} [H]
        \centering
        \includegraphics[width=0.8\linewidth]{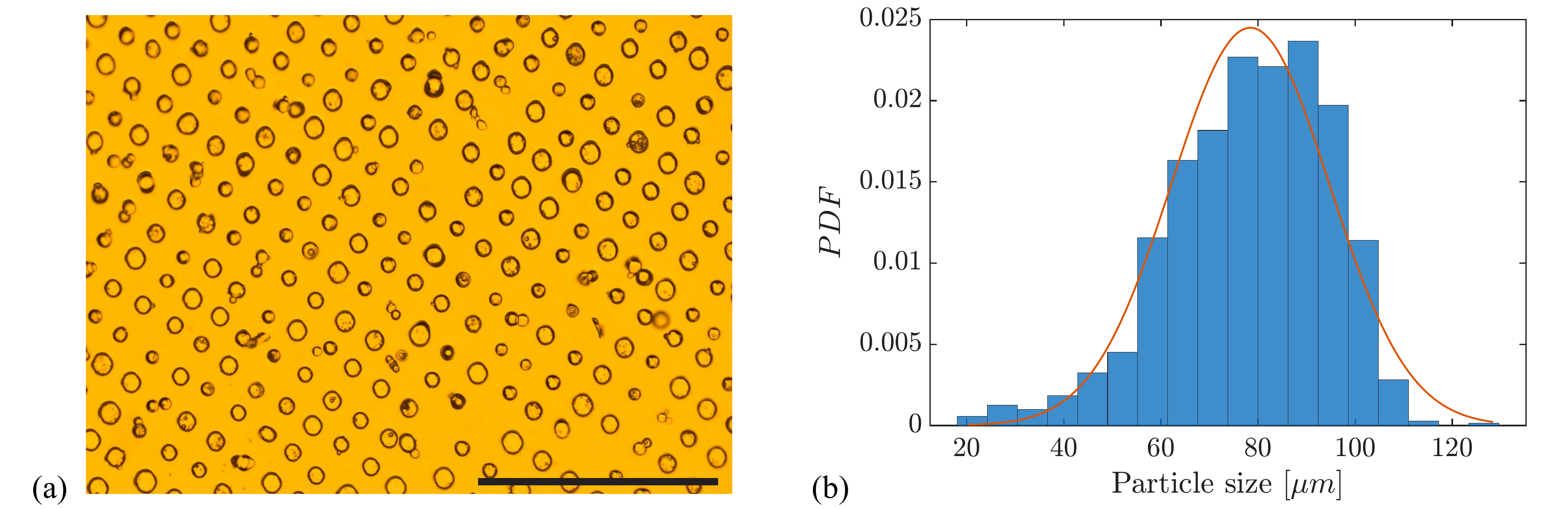}
        \caption{(a) Representative microscopic image of the particles used in the experiments. Scale bar: 1 mm. (b) Particle size distribution obtained from image analysis. The solid line shows a Gaussian fit. }
        \label{fig:SI1}
    \end{figure}

Figure~\ref{fig:SI1}(a) shows a representative image of the particles used in the experiments obtained using an optical microscope. The images were then analyzed with ImageJ. The projected area $A$ of each particle was measured, and the equivalent circular diameter was calculated as
\begin{equation*}
d=\sqrt{\frac{4A}{\pi}} .
\end{equation*}

\noindent The resulting particle-size distribution, shown in Fig.~\ref{fig:SI1}(b), can be described by a Gaussian fit, with a mean particle diameter of $78\,\mu\mathrm{m}$ and a standard deviation of $16\,\mu\mathrm{m}$.

\bigskip
\clearpage


\section*{Initial drop diameter and impact velocity}

\smallskip

\begin{figure} [H]
        \centering
        \includegraphics[width=\linewidth]{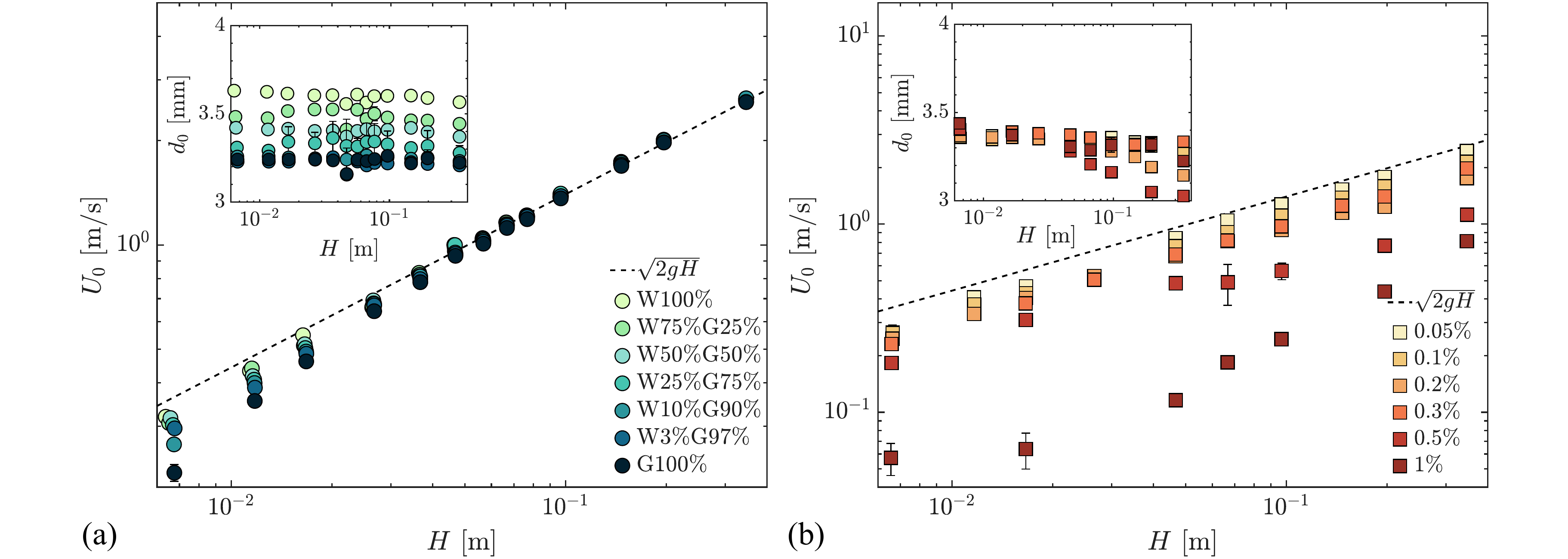}
        \caption{Measured impact velocity $U_0$ of (a) Newtonian and (b) PEO solutions as a function of release height $H$. The dashed line indicates the nominal free-fall prediction $U_{\rm ff}=\sqrt{2gH}$. Insets show the corresponding initial drop diameter $d_0$.}
        \label{fig:SI2}
    \end{figure}

Figure~\ref{fig:SI2} reports the measured impact velocity $U_0$ and initial drop diameter $d_0$ for each release height $H$. The dashed line corresponds to the free-fall prediction, $U_{\rm ff}=\sqrt{2gH}$, for a drop released from rest over a distance $H$. For the Newtonian glycerol--water mixtures [Fig.~\ref{fig:SI2}(a)], the measured impact velocity follows the free-fall prediction over most of the range of release heights. However, small deviations are observed, especially at the lowest release heights and for the mixture with a large fraction in glycerol. These deviations at small heights are due to finite drop size and pinch-off, since the actual drop center of mass at pinch-off is not exactly located at the needle tip. The inset shows that, for a given Newtonian liquid, $d_0$ remains nearly independent of $H$, although it varies slightly with liquid composition.

\medskip

\noindent The behavior is different for the PEO solutions [Fig.~\ref{fig:SI2}(b)]. In this case, the measured impact velocity is generally lower than the free-fall prediction and shows larger variations, especially at higher polymer concentrations. This difference is consistent with the formation of a viscoelastic filament between the needle and the falling drop during detachment. The filament can keep the drop mechanically coupled to the needle over part of the fall and can reduce the effective impact velocity. The inset further shows that $d_0$ decreases with increasing release height and polymer concentration, most clearly for the $1\%$ PEO solution. This trend is consistent with part of the liquid remaining in the stretched filament rather than in the impacting drop.

\medskip

\noindent These measurements show that the release height $H$ alone is not sufficient to define the impact conditions, especially for the polymer solutions. Therefore, all dimensionless groups and impact energies reported in the main text are calculated using the measured values of $U_0$ and $d_0$ for each experiment.

\bigskip
\clearpage




\section*{Rheology of the PEO solutions}

\begin{figure} [H]
        \centering
        \includegraphics[width=0.5\linewidth,pagebox=cropbox]{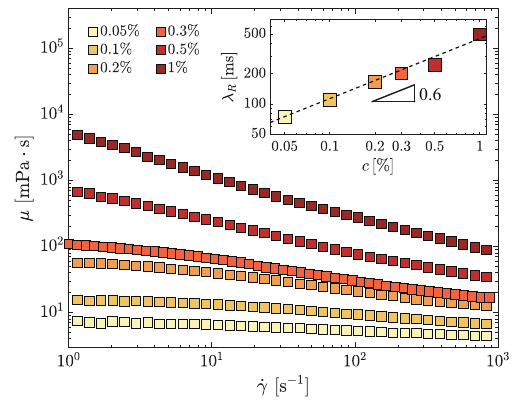}
        \caption{Dynamic viscosity $\mu$ of 4M PEO solutions in a 75/25 wt\% water/glycerol solvent as a function of shear rate $\dot{\gamma}$. Inset: relaxation time $\lambda_{\rm R}$ as a function of polymer concentration $c$. The dashed line is a power-law guide with exponent $0.6$.}
        \label{fig:SM_rheology}
    \end{figure}

\noindent Figure~\ref{fig:SM_rheology} shows the shear-rate-dependent viscosity of the PEO solutions used in the experiments. The inset reports the corresponding relaxation time $\lambda_{\rm R}$ as a function of polymer concentration. These rheological measurements are used to determine the effective viscosity entering the dimensionless groups in the main text.

\bigskip

\clearpage


\section*{Effective viscosity of PEO}

\noindent The PEO solutions are shear-thinning, so the viscosity relevant during impact must be evaluated at a representative shear rate. We estimate this shear rate using the characteristic shear rate framework of Mobaseri \textit{et al.}, developed for the maximum spreading of impacting generalized Newtonian drops.~\cite{Mobaseri2025PNAS} In this framework, the characteristic shear rate is written as $\dot{\gamma}=U_0/\delta$, where $\delta$ is an effective viscous length scale associated with lamella spreading. Physically, $\delta$ represents an effective viscous thickness selected by a boundary-layer-controlled mechanism during impact. In the smooth-substrate framework of Mobaseri \textit{et al.}, this length scale results from the competition between the geometric thinning of the lamella and the growth of a viscous boundary layer within the spreading drop. Therefore, $\delta$ provides a more appropriate length scale for estimating the dominant shear rate during spreading than the purely geometric minimum thickness.

\smallskip

\noindent The characteristic viscous length scale $\delta$ depends on the Reynolds number according to the scaling relation proposed by Mobaseri \textit{et al.},~\cite{Mobaseri2025PNAS}
\begin{equation}
\frac{\delta}{d_0} \simeq \Rey^{-0.4 + 0.06 \exp(-0.03\We)}.
\end{equation}
Here, the prefactor is close to unity and is taken as 1. In the present work, this relation is used as a practical estimate of the shear rate controlling the effective shear viscosity during spreading. It is not intended to describe the full viscoelastic response of the PEO solutions or the complete liquid--grain interaction on the granular bed. Elastic effects are characterized separately through the Deborah number defined in the main text.

\smallskip

\noindent Because $\delta$ depends on $\Rey$, and $\Rey$ itself depends on the effective viscosity evaluated at $\dot{\gamma}_{\rm c}=U_0/\delta$, the quantities $\delta$, $\dot{\gamma}_{\rm c}$, $\mu_{\rm eff}$, and $\Rey$ must be determined self-consistently. As an initial estimate, we first compute the geometric minimum lamella thickness at maximum spreading, $h_{\min} = 2 {d_0}^3/(3\,{d_{\max}}^2)$, where $d_0$ is the initial droplet diameter and $d_{\max}$ is the maximum spreading diameter. This gives an initial shear rate $\dot{\gamma}_0=U_0/h_{\min}$, from which an initial viscosity $\mu_0$ is obtained using the measured rheological curve of the PEO solution.

\smallskip

\noindent Starting from $\mu_0$, we compute $\Rey$, update $\delta$ using the scaling above, calculate a new shear rate $\dot{\gamma}_{\rm c}=U_0/\delta$, and obtain a new viscosity from the measured rheology. This procedure is repeated until the relative change in $\delta$ between successive iterations is smaller than $1\%$. The converged viscosity is then used as the effective viscosity, $\mu_{\rm eff}$, to calculate $\Rey$ and $\Oh$ for each PEO impact condition.

\smallskip

\begin{figure}[H]
        \centering
        \includegraphics[width=\linewidth]{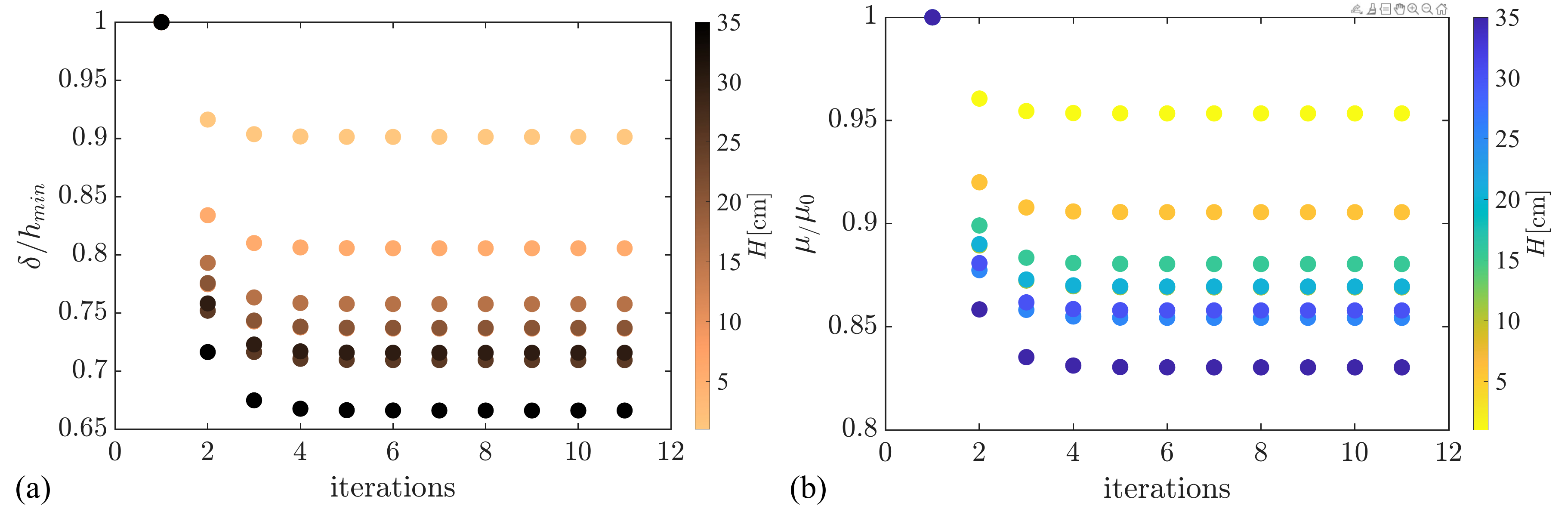}
        \caption{(a) Evolution of the normalized characteristic viscous length scale, $\delta/h_{\min}$, with iteration number. (b) Evolution of the normalized effective viscosity, $\mu_{\rm eff}/\mu_0$, with iteration number. The color scale denotes the release height $H$.}
        \label{fig:SI3}
\end{figure}

\noindent Figure~\ref{fig:SI3} shows the convergence of $\delta/h_{\min}$ and $\mu_{\rm eff}/\mu_0$ with iteration number. Both quantities converge within a few iterations, indicating that the procedure is stable. As the release height increases, the characteristic shear rate increases and the effective viscosity decreases, consistent with the shear-thinning behavior of the PEO solutions.

\clearpage

\begin{figure} [H]
        \centering
        \includegraphics[height=0.9\textheight]{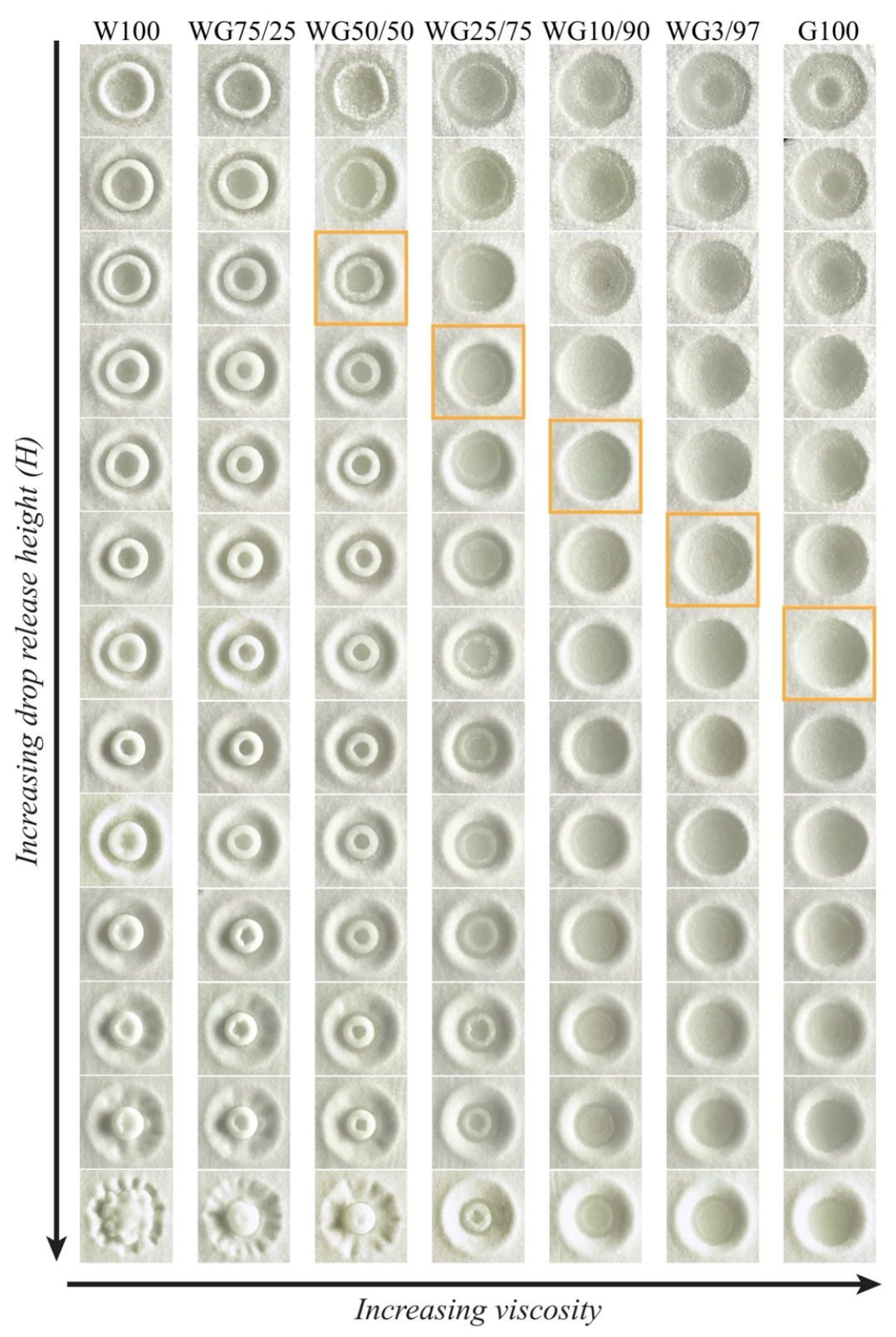}
        \caption{Crater image map for Newtonian solutions. The orange boxes denote the transition point between the plateau and the power-law crater growth region. (Not to scale)}
        \label{fig:SI4}
    \end{figure} 
\clearpage

\begin{figure}[H]
        \centering
        \includegraphics[height=0.9\textheight]{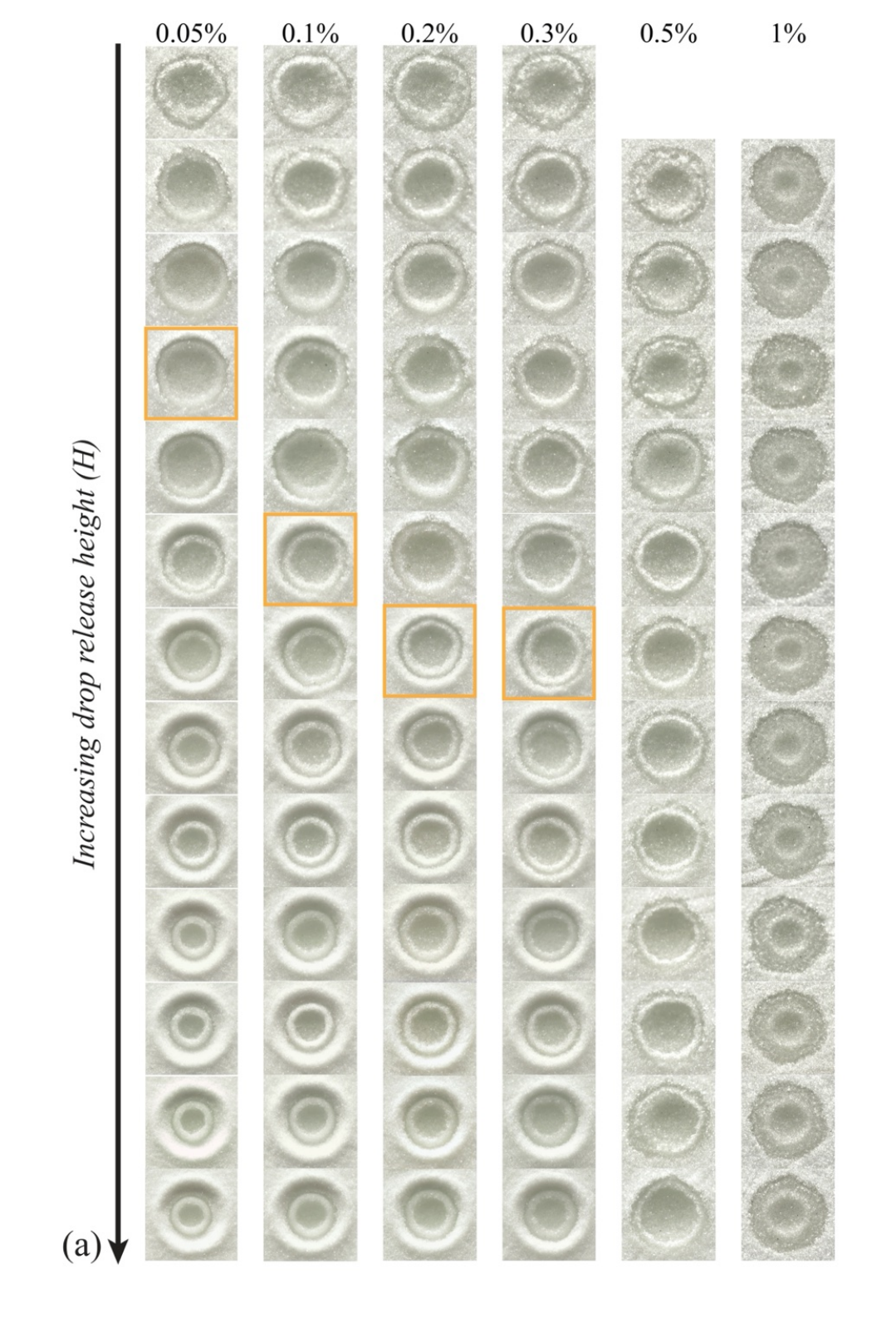}
        \caption{Crater image map for viscoelastic (4M PEO) solutions at lower drop release heights. The orange boxes denote the transition point between the plateau and the power-law crater growth region. Images are cropped individually for visibility and are not to scale.}
        \label{fig:SI5a}
\end{figure}

\begin{figure}[H]
        \centering
        \includegraphics[height=0.67\textheight]{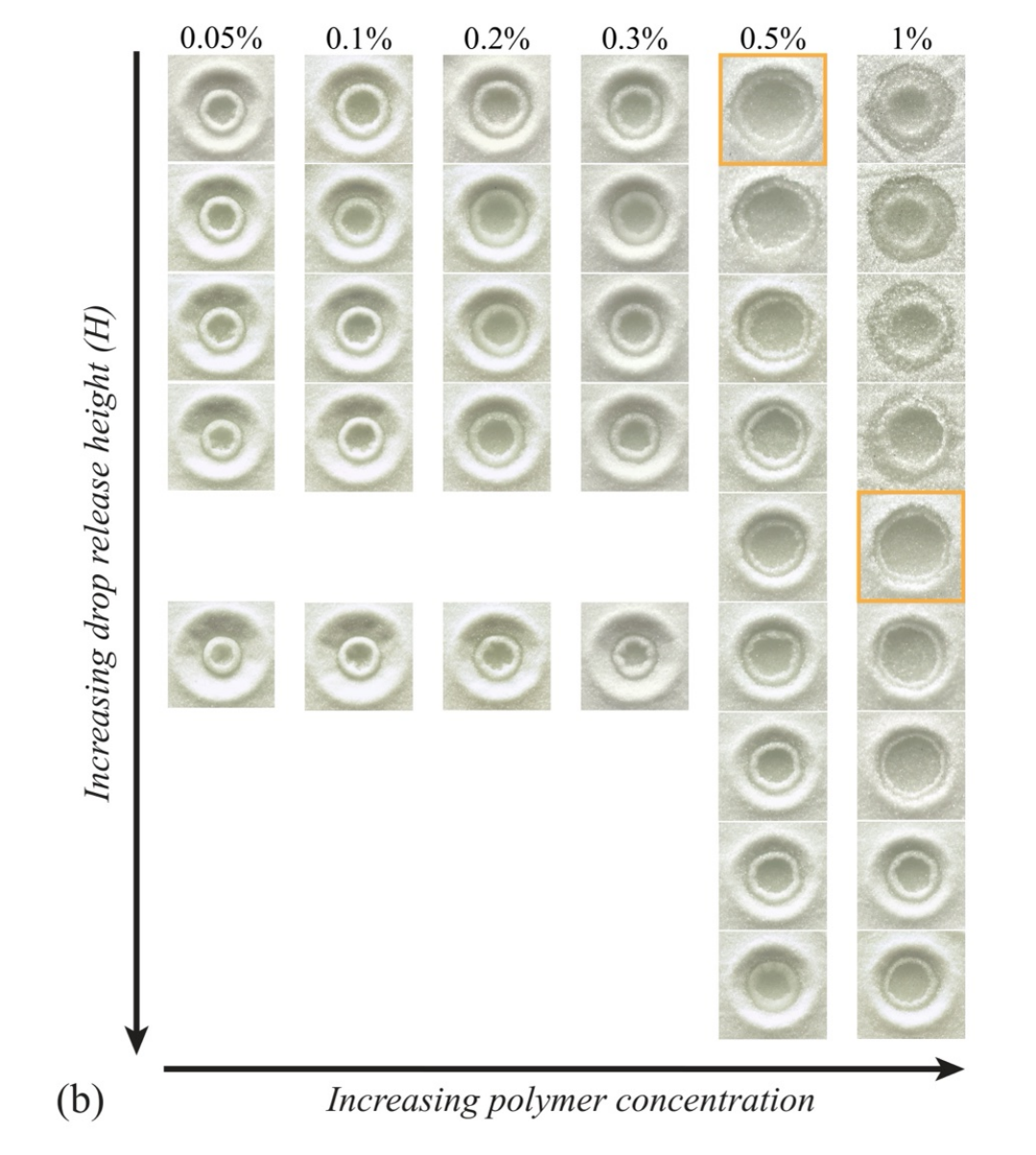}
        \caption{Crater image map for viscoelastic (4M PEO) solutions at higher drop release heights. The orange boxes denote the transition point between the plateau and the power-law crater growth region. Images are cropped individually for visibility and are not to scale.}
        \label{fig:SI5b}
\end{figure}
\clearpage

\section*{Crater Morphologies: Bowl and Stain}

\begin{figure} [H]
        \centering
        \includegraphics[width=\linewidth]{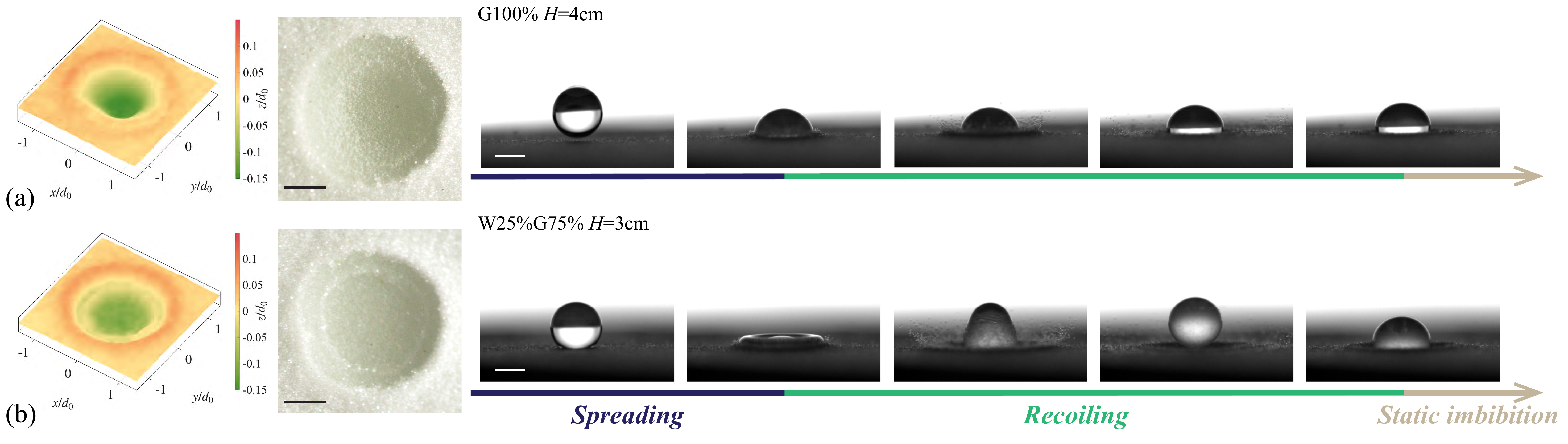}
        \caption{Representative crater morphologies for Newtonian drops. (a) Deep bowl morphology produced by a highly viscous Newtonian liquid ($\Oh>1$: G100\%, $H=4~\mathrm{cm}$). (b) Stain morphology observed for a lower-viscosity Newtonian liquid ($\Oh<1$: W25\%G75\%, $H=3~\mathrm{cm}$). Scale bars: 2~mm.}
        \label{fig:SI6}
    \end{figure}

In the main text, the bowl and stain morphologies were discussed mainly for fluids with lower Ohnesorge numbers ($\Oh<1$), for which the crater boundary and inner morphology are relatively easy to identify. For Newtonian fluids with $\Oh>1$, the resulting crater shapes are generally deeper. In the present study, the $\Oh>1$ cases correspond to highly viscous Newtonian liquids (W3\%G97\% and G100\%). These fluids exhibit much weaker spreading behavior, with the normalized maximum spreading diameter, $\bar{d}_{\max}=d_{\max}/d_0$, remaining nearly constant as the Weber number $\We$ increases. As a result, compared with the $\Oh<1$ fluids, the deformation of the granular bed is concentrated closer to the drop axis while the drop undergoes limited radial spreading (Fig.~\ref{fig:SI6}(a)). This leads to the formation of a deeper crater with a relatively steep wall after impact. Grains on this steep slope can slide downward, making it difficult to clearly distinguish the boundary corresponding to the early drop impact footprint. Consequently, the bowl morphology produced by these high-viscosity Newtonian fluids differs from that observed for the PEO solutions.

\medskip

\noindent Even within the $\Oh<1$ regime, the stain morphology also differs between Newtonian and PEO liquids. For PEO solutions, the stain morphology forms as a broad, circular deposit of particles near the outer part of the bowl. In contrast, for Newtonian liquids, the stain typically appears as a thin band inside the bowl, and the morphology evolves toward the formation of this inner ring (Fig.~\ref{fig:SI6}(b)). The physical origin of this difference is discussed in the main text.

\vspace{1.5cm}


\section*{Determination of the transition energy}

For each fluid, the transition energy $E_{\rm t}$ was determined from the evolution of the normalized crater diameter $\bar{D}_c$ with the impact kinetic energy $E_{\rm k}$. The low-energy data were fitted by an energy-independent plateau, $\bar{D}_{c,0}$, while the high-energy data were fitted by a power law, $\bar{D}_c=A {E_{\rm k}}^{\alpha}$. Here, $A$ and $\alpha$ are fitting parameters determined independently for each fluid. The transition energy was then defined as the intersection between these two fitted curves:
\begin{equation*}
E_t=\left(\frac{\bar{D}_{\rm c,0}}{A}\right)^{1/\alpha}.
\end{equation*}
This procedure provides a quantitative estimate of the onset of crater growth, rather than assigning the transition from a single image. Since the fitted value of $E_t$ does not necessarily coincide exactly with one of the experimental impact energies, the orange boxes in Figs.~\ref{fig:SI4}, \ref{fig:SI5a}, and \ref{fig:SI5b} indicate the crater image measured at the impact energy closest to $E_t$ for each fluid.

\vspace{2cm}

\section*{Post-recoil imbibition timescale}

The post-recoil imbibition timescale $t_{\rm imb}$ was measured from the impact experiments at the transition condition. For each liquid, we analyzed the high-speed videos for the experiment performed at the impact energy closest to the fitted transition energy $E_t$. This timescale characterizes the late-absorption stage following spreading and recoil. It should therefore not be interpreted as a direct measurement of the fast pressure-driven penetration that may occur during the drop impact.

\medskip

\noindent For each liquid, the start of the imbibition stage was defined as the first frame in which the droplet remained in contact with the bed and no longer exhibited noticeable macroscopic spreading or recoil. The end of imbibition was defined as the first frame in which no distinct liquid cap remained visible above the granular surface. The difference between these two times leads to the imbibition timescale $t_{\rm imb}$.

\medskip

\noindent The impact timescale was estimated as $t_{\rm imp}=d_0/U_0$, using the measured drop diameter and impact velocity at the transition point. The ratio $t_{\rm imp}/t_{\rm imb}$ reported in the main text was computed for each liquid at this transition condition. 

\vspace{2cm}

\section*{Plateau size and power-law exponents}

\begin{figure} [H]
        \centering
        \includegraphics[width=\linewidth]{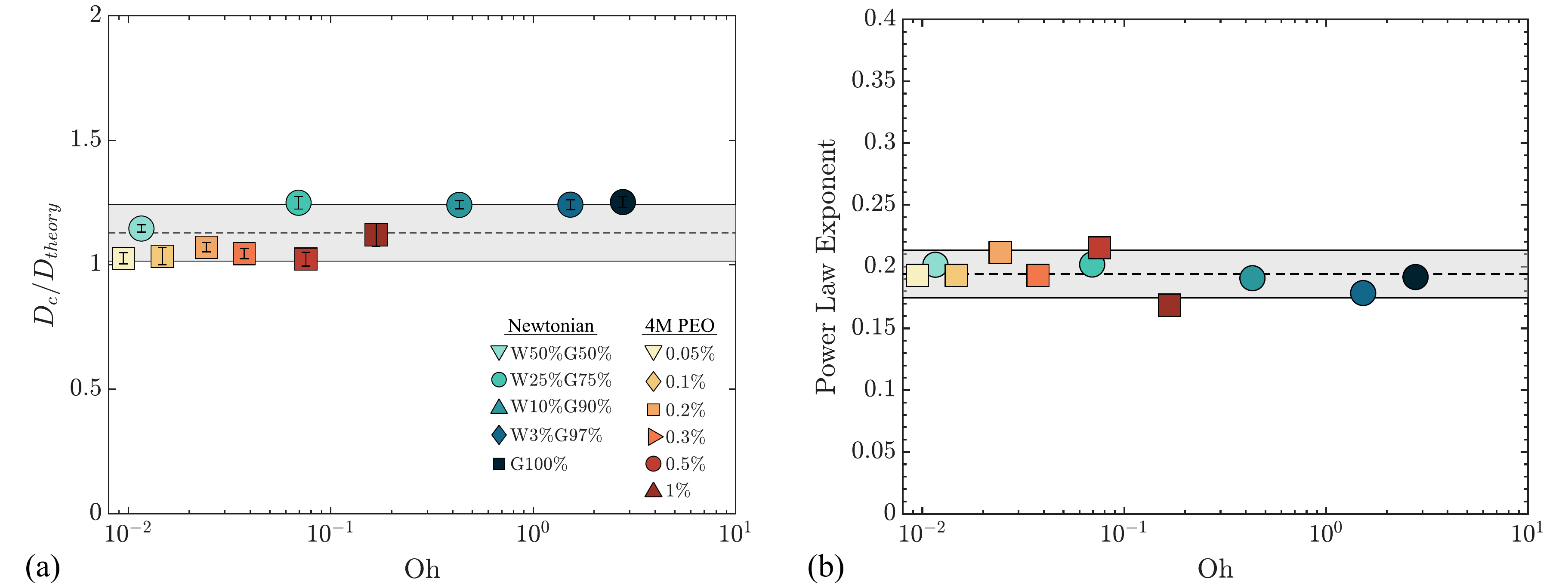}
        \caption{(a) Plateau crater diameter normalized by the theoretical estimate, $D_c/D_{\rm theory}$, and (b) power-law exponents as a function of $\Oh$. The gray shaded bands denote the $\pm10\%$ range around the average value.}
        \label{fig:SI7}
    \end{figure}

Figure~\ref{fig:SI7}(a) shows the plateau crater diameter normalized by the theoretical estimate, $D_c/D_{\rm theory}$, for all fluids as a function of $\Oh$. A volume-conservation estimate yields $D_{\rm theory} = \left( 2/{(1-\phi)} \right)^{1/3} d_0$, where $\phi$ is the packing fraction of the granular bed. All cases fall within $\pm10\%$ of the mean value, indicating that $D_c$ is consistently about $1.1\,D_{\rm theory}$ and shows no clear dependence on $\Oh$. 

\medskip

\noindent Figure~\ref{fig:SI7}(b) shows the power-law exponent for all fluids as a function of $\Oh$. The average exponent is $0.194$, and all data lie within $\pm10\%$ of this value. This range is consistent with previous studies reporting exponents close to $0.17-0.25$, reflecting the combined influence of drop spreading and granular-bed deformation on crater formation.

\vspace{2cm}

\section*{Maximum drop spreading diameter at the transition point}

\begin{figure} [H]
        \centering
        \includegraphics[width=0.5\linewidth]{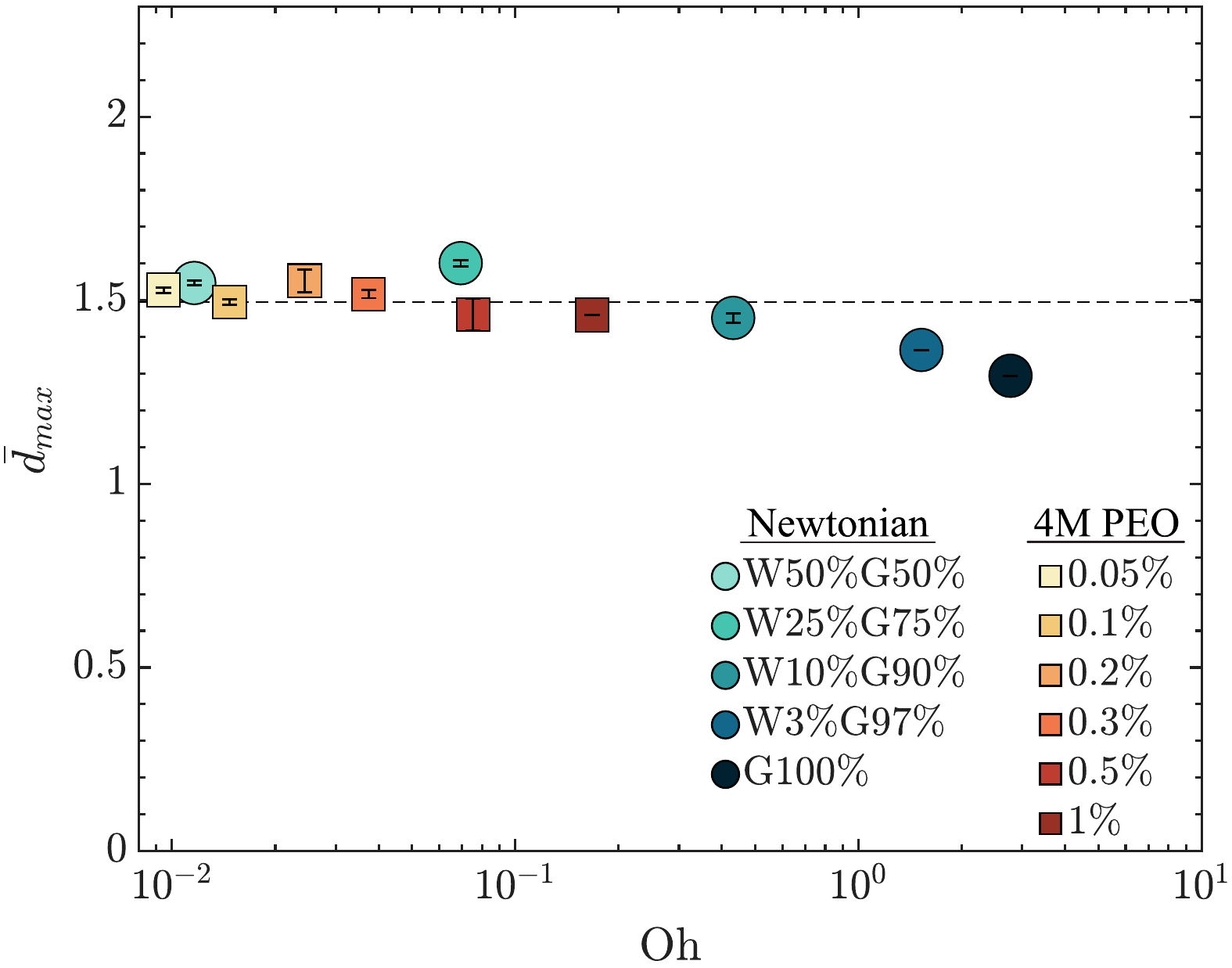}
        \caption{Normalized maximum drop spreading diameter, $\bar{d}_{\max}=d_{\max}/d_0$, at the transition points as a function of $\Oh$. The black dashed line indicates the average value, $\bar{d}_{\max} \approx 1.5$.}
        \label{fig:discussion3}
    \end{figure}

Figure~\ref{fig:discussion3} shows the normalized maximum drop spreading diameter, $\bar{d}_{\max}=d_{\max}/d_0$, at the transition points for all fluids. Remarkably, $\bar{d}_{\max}$ converges to approximately 1.5, with only a weak dependence on fluid type and $\Oh$. Since the impact energy required to reach the transition point is lower for PEO solutions, this suggests differences in how the impact energy is partitioned during impact.

\vspace{2cm}

\end{document}